\PassOptionsToPackage{pdfpagelabels=false}{hyperref}
\documentclass[fleqn,usenatbib]{mnras}
\usepackage{xcolor, graphicx}
\usepackage{amsmath}
\usepackage{amssymb}
\usepackage{soul}
\usepackage{txfonts}
\usepackage{journal_names}

\newcommand{\Msun}{\rm M_\odot}
\newcommand{\kms}{\mbox{km s$^{-1}$}}
\newcommand{\SBunit}{\mbox{mag arcsec$^{-2}$}}
\newcommand{\re}{R_{\rm e}}

\def\degr{\hbox{$^\circ$}}

\def\arcsec{\hbox{$^{\prime\prime}$}}

\def\farcs{\hbox{$.\!\!^{\prime\prime}$}}

\def\Sersic/{{S\'ersic}}

\title[Origins of ultra-diffuse galaxies in the Coma cluster -- I.]{Origins of ultra-diffuse galaxies in the Coma cluster -- I. Constraints from velocity phase-space}

\author[Alabi et al. ]{Adebusola Alabi$^{1}$\thanks{E-mail: aalabi@ucsc.edu }, Anna Ferr\'e-Mateu$^{2}$, Aaron J. Romanowsky$^{1,3}$, Jean Brodie$^{1}$,  
\newauthor  Duncan A. Forbes$^{2}$, Asher Wasserman$^{1}$, Sabine Bellstedt$^{2}$, Ignacio Mart\'in-Navarro$^{1}$,   
\newauthor  Viraj Pandya$^{1}$, Maria Stone$^{3}$, Nobuhiro Okabe$^{4,5,6}$\\
$^{1}$ University of California Observatories, 1156 High St., Santa Cruz, CA 95064, USA\\
$^{2}$Centre for Astrophysics \& Supercomputing, Swinburne University of Technology, Hawthorn VIC 3122, Australia\\
$^{3}$ Department of Physics and Astronomy, San Jos\'e State University, San Jose, CA 95192, USA \\
$^{4}$ Department of Physical Science, Hiroshima University, 1-3-1, Kagamiyama, Higashi-Hiroshima, Hiroshima 739-8526, Japan\\
$^{5}$ Hiroshima Astrophysical Science Center, Hiroshima University, 1-3-1, Kagamiyama, Higashi-Hiroshima, Hiroshima 739-8526, Japan\\
$^{6}$ Core Research for Energetic Universe, Hiroshima University, 1-3-1, Kagamiyama, Higashi-Hiroshima, Hiroshima 739-8526, Japan\\
}

\begin{document}

\date{Submitted to MNRAS; Accepted XXX. Received YYY; in original form ZZZ}
\pubyear{2018}

\pagerange{\pageref{firstpage}--\pageref{lastpage}} \pubyear{2018}

\maketitle

\label{firstpage}
\begin{abstract} We use Keck/DEIMOS spectroscopy to confirm the cluster membership of 16 ultra-diffuse galaxies (UDGs) in the Coma cluster, bringing the total number of spectroscopically confirmed UDGs to 24. We also identify a new cluster background UDG. In this pilot study of Coma UDGs in velocity phase-space, we find evidence that most present-day Coma UDGs have a recent infall epoch while a few may be ancient infalls. These recent infall UDGs have higher absolute relative line-of-sight velocities, bluer optical colors, and are smaller in size, unlike the ancient infalls. The kinematics of the spectroscopically confirmed Coma UDG sample is similar to that of the cluster late-type galaxy population. Our velocity phase-space analysis suggests that present-day cluster UDGs have a predominantly accretion origin from the field, acquire velocities corresponding to the mass of the cluster at accretion as they are accelerated towards the cluster center, and become redder and bigger as they experience the various physical processes at work within the cluster.
\\
\\
\end{abstract}


\section{Introduction}
One of the overarching challenges that has attended the discovery of ultra-diffuse galaxies (UDGs) is properly reconciling their observed properties with the various scenarios that have been advanced for their formation. UDGs, which have now been found in diverse environments (e.g. \citealt{vanDokkum_2015, vdBurg_2016, vdBurg_2017, Janssens_2017, Roman_2017b}), have low surface brightnesses ($\geq24\ \SBunit$), sizes that are comparable to or even larger than $L^*$ ellipticals, luminosities consistent with dwarf galaxies (${\sim}10^{8}\ \rm L_\odot$), but in some cases associated with massive halos. They have also been successfully identified in cosmological simulations \citep[e.g.][]{Rong_2017, Chan_2017, DiCintio_2017}. The two major scenarios that attempt to explain the origin of UDGs describe them either as failed galaxies or puffy dwarfs, with both emerging paradigms increasingly finding support from observations.

For example, initial studies of the spatial distribution (e.g., \citealt{vdBurg_2016, Roman_2017a}), colors (e.g., \citealt{Beasley_2016, Yagi_2016}), and the shapes (\citealt{Burkert_2017}) of UDGs in dense environments point at a similarity to dwarfs, while the formation landscape is ill-defined when one considers the central velocity dispersion \citep{vanDokkum_2016}, the stellar populations, and the globular cluster systems in the handful of UDGs so far studied. It appears that UDGs in dense environments may be dominated by intermediate-to-old and metal poor stellar populations (e.g., \citealt{Meng_2017, Viraj_2017, Kadowaki_2017}), while in less-dense environments, some UDGs have been shown to host younger stellar populations (e.g., DGSAT~I \citealt{Viraj_2017} and the UDGs in the Hickson Compact groups \citealt{Shi_2017, Spekkens_2017}, respectively), with hints of an extended star formation history. Results from complementary studies of the globular cluster systems associated with UDGs reveal populations that are indicative of either dwarfs or $L^*$ galaxies (\citealt{Beasley_2016}: DF17 with 27 GCs; see also \citealt{Amorisco_2016}, \citealt{vanDokkum_2017}: DF44 and DFX1 with ${\sim}74$ and ${\sim}62$ GCs).

\begin{figure*}
    \includegraphics[scale=0.8]{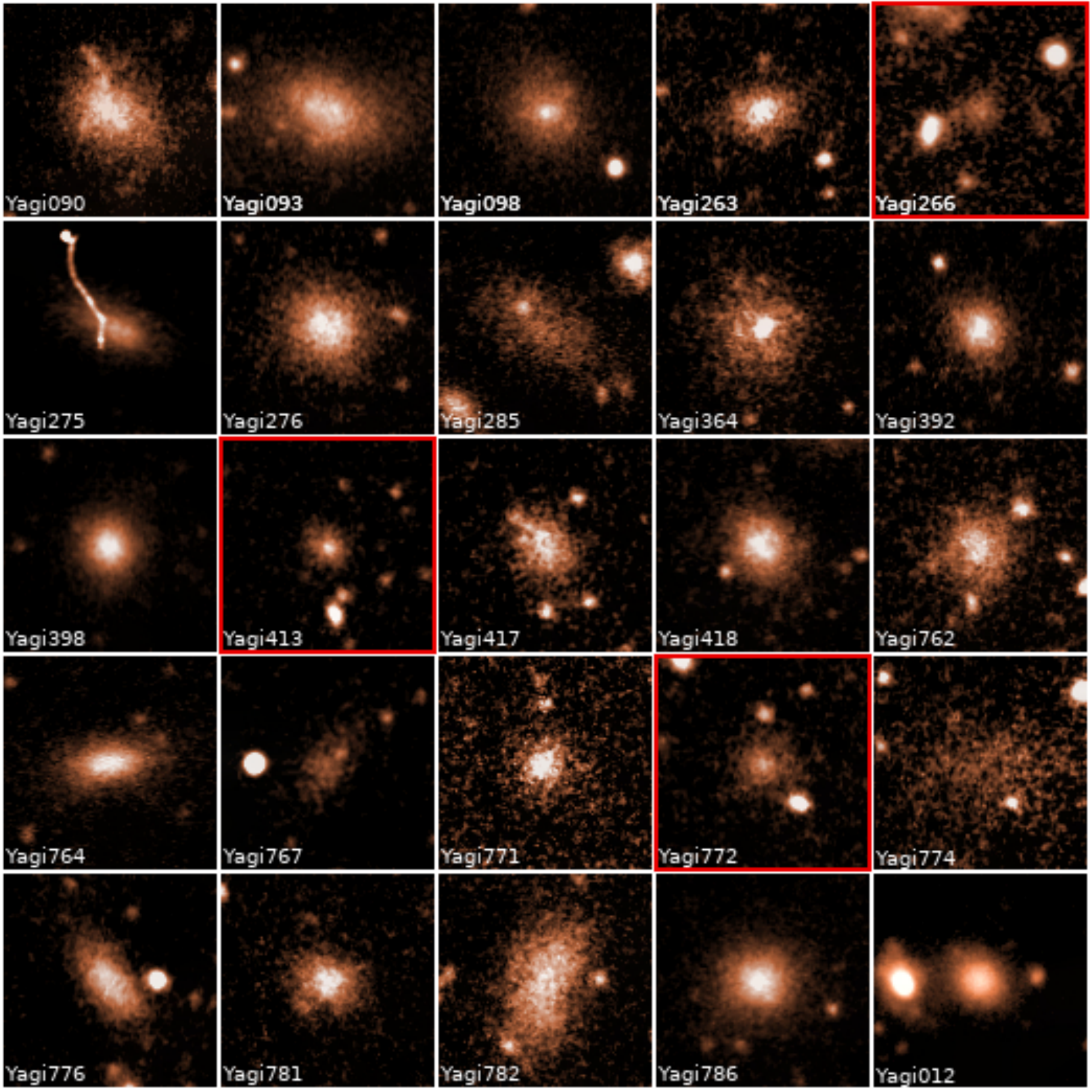}\hspace{0.01\textwidth}\\
	\caption{\label{fig:sample_img} Thumbnails of the 25 Coma cluster low surface brightness (LSB) galaxies studied in this work, from $V$-band Suprime-Cam/Subaru imaging. Each thumbnail is $10\times10$~kpc across. North is up and east is to the left. Three of our LSB galaxies fail the $\re > 1.5$~kpc criterion from \citet{vanDokkum_2015} and they are no ultra-diffuse galaxies. These galaxies have been highlighted in the diagram with red borders. The galaxy IDs shown here are from the Coma LSB catalog of \citet{Yagi_2016}. We report spectroscopic redshifts for 16 new Coma ultra-diffuse galaxies and a Coma cluster background galaxy in this work. The stream-like feature super-imposed on Yagi~275 is an image flaw.}
\end{figure*}

It may be possible to gain clearer insights into the origin of UDGs by studying them in velocity phase-space \citep{Bertschinger_1985, Mamon_2004, Mahajan_2011, Oman_2013, VijarX_2015, Haines_2015, Rhee_2017}.
\citet{Conselice_2001} used the velocity phase-space analysis to show the late accretion origin of the dwarf galaxy population in the Virgo cluster, relative to the cluster giants. Recently accreted groups of galaxies that are yet to be completely disrupted within the cluster potential may still retain a memory of their origins, showing up as ``lumps'' or ``streams'' around some giant galaxies. \citet{Mendelin_2017} recently found a significant excess of faint galaxies around giant spiral galaxies in the Coma cluster, mostly at large clustercentric radii, and explained their result as evidence of the hierarchical buildup of the cluster.

For a Coma-sized cluster, where the relaxation and energy equipartition timescales are significantly longer than the Hubble time ($> 2.63 \times 10^{11}$~Gyr, using eqn. 7.1 from \citet{Binney_2008} and ${\sim}22$~Gyr, considering only the most massive galaxies and using eqn. 2.36 from \citealt{Sarazin_1986}, respectively), it is reasonable to explore phase-space for evidence in support of or against the recent infall hypothesis. To first order, galaxies populate distinct phase-space regions according to their accretion epochs, with their radial velocities reflecting the cluster mass at infall. UDGs (and other cluster galaxies) accreted at earlier epochs should be virialized. These galaxies would have experienced the quenching effects of the various physical processes operating within the cluster environments for a longer time, and they should have passively evolved to be maximally red in optical colors. Recent infalling UDGs, on the other hand, are expected to have higher velocities, reflecting the cluster mass at infall, and to be bluer (in a standard dwarf-like picture).

\begin{figure}
    \includegraphics[width=0.48\textwidth]{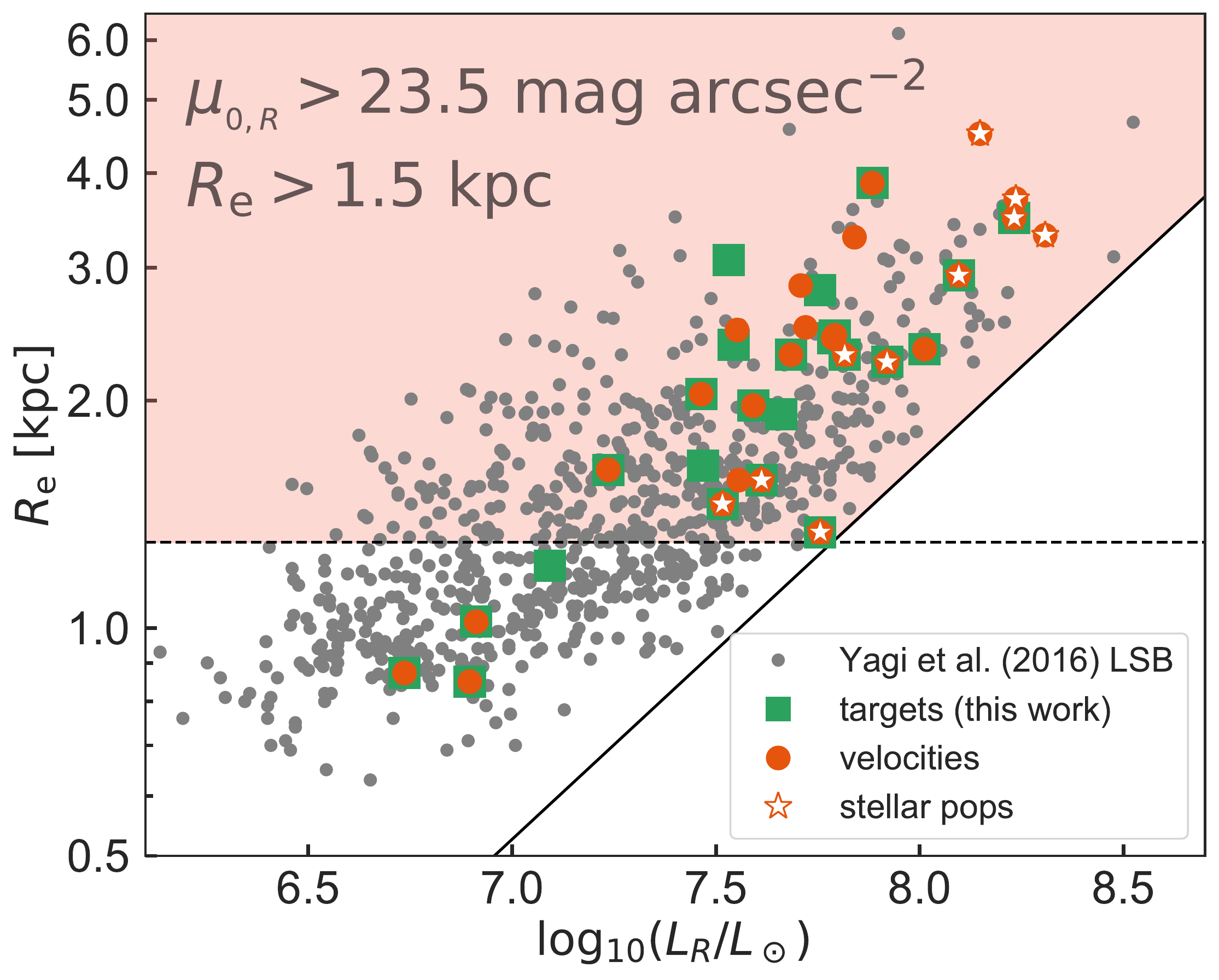}\hspace{0.01\textwidth}\\
	\caption{\label{fig:sample_fig} Size--luminosity diagram of our low surface brightness (LSB) spectroscopic galaxy sample. We have marked the 25 targets studied in this work (green squares), chosen from the \citet{Yagi_2016} LSB catalog (gray dots). Galaxies with radial velocities from this work and in the literature as well as those with stellar population parameters from F\'erre-Mateu et al. (2018), submitted, and in the literature, have been marked as shown in the plot legend. The shaded region, defined by the dashed line, which corresponds to the $\re > 1.5$~kpc criterion from \citet{vanDokkum_2015}, have been scaled to reflect that sizes from \citet{Yagi_2016} are 15 per cent smaller than those of \citet{vanDokkum_2015}, and the slanted solid line, which is a line of constant mean surface brightness ($\mu_{_{0}}=23.5~\SBunit$ in $R$-band), show that most of our targets are consistent with the UDG definitions in the literature.}
\end{figure}

Phase-space exploration of UDGs is however challenging and expensive primarily due to their faintness. Even in a UDG-rich environment, such as the Coma cluster, which may host ${\sim}200-300$~UDGs (\hypertarget{Y16}{\citealt{Yagi_2016}}, hereafter \hyperlink{Y16}{Y16}, and \citealt{Janssens_2017}), only 8 of them have been spectroscopically confirmed \citep{vanDokkum_2016, Kadowaki_2017, Meng_2017, vanDokkum_2017} from a combined $\geq 54$~hrs of observation, mostly with $10$-m  class telescopes. Apart from the intrinsic faintness of UDGs, the relative proximity of the Coma cluster at ${\sim}100$~Mpc, and thus its angular extent of $\geq 2 \degr$ on the sky, makes an attempt to obtain a large representative sample of UDG radial velocities extremely time consuming. Nevertheless, in this paper, we report the spectroscopic confirmation of 18 new UDGs in the Coma cluster and test the recent infall hypothesis by comparing UDG kinematics with other cluster galaxy populations. This paper is the first in a series based on new Keck/DEIMOS data that seek to understand the origins of UDGs in cluster environments (see F\'erre-Mateu et al. 2018 (submitted), hereafter Paper II, where we study the stellar populations of a large sample of Coma UDGs to better understand their origins).

The outline of the paper is as follows: Section \ref{data} describes our sample selection, observational setup, and data reduction methodology. In Section \ref{analysis}, we present our radial velocity measurements. In Section \ref{discussion}, we use our results to address some fundamental questions about the origins of UDGs in cluster environments. Throughout this work, we adopt a distance of 100~Mpc, a virial radius of ${\sim}2.9$~Mpc, and a virial mass of ${\sim}2.7 \times 10^{15}\ \Msun$ for the Coma cluster \citep{Kubo_2007}. We use a mean heliocentric radial velocity of $6943\ \kms$, a central velocity dispersion of $1031\ \kms$, and central co-ordinates RA: 12:59:48.75 and Dec: +27:58:50.9 for the Coma cluster \citep{Hyperleda}. Lastly, we adopt the following cosmology: $\rm \Omega_m=0.3$, $\rm \Omega_{\Lambda}=0.7$ and, $\rm H_{0}=70\ \kms\ Mpc^{-1}$.

\section{Data: Sample Selection, Observations and Reduction}
\label{data}
\subsection{Sample Selection}
\label{sample_sel}
We obtained spectroscopic data for a sample of Coma cluster low surface brightness (LSB) galaxies from the Subaru-LSB catalog of \hyperlink{Y16}{Y16} (see Figure \ref{fig:sample_img} for a montage of our LSB sample). It should be noted that most of the LSB galaxies from \hyperlink{Y16}{Y16} fall short of the UDG definition in \citet{vanDokkum_2015}, i.e., $\re > 1.5$~kpc and $\mu_{_{0}} > 24~\SBunit$, in the $g$~band (equivalent to $23.5~\SBunit$ in the $R$-band). Therefore, we select 25 LSB galaxies from the \hyperlink{Y16}{Y16} catalog that maximize the number of unambiguous UDGs and simultaneously include targets from the core and outskirts regions of the cluster. Our sample has 6 targets in common with the \citet{vanDokkum_2015} catalog: Yagi093 (DF26), Yagi276 (DF28), Yagi285 (DF25), Yagi364 (DF23), Yagi762 (DF36), and Yagi782 (DF32). 

\begin{figure}
    \includegraphics[width=0.48\textwidth]{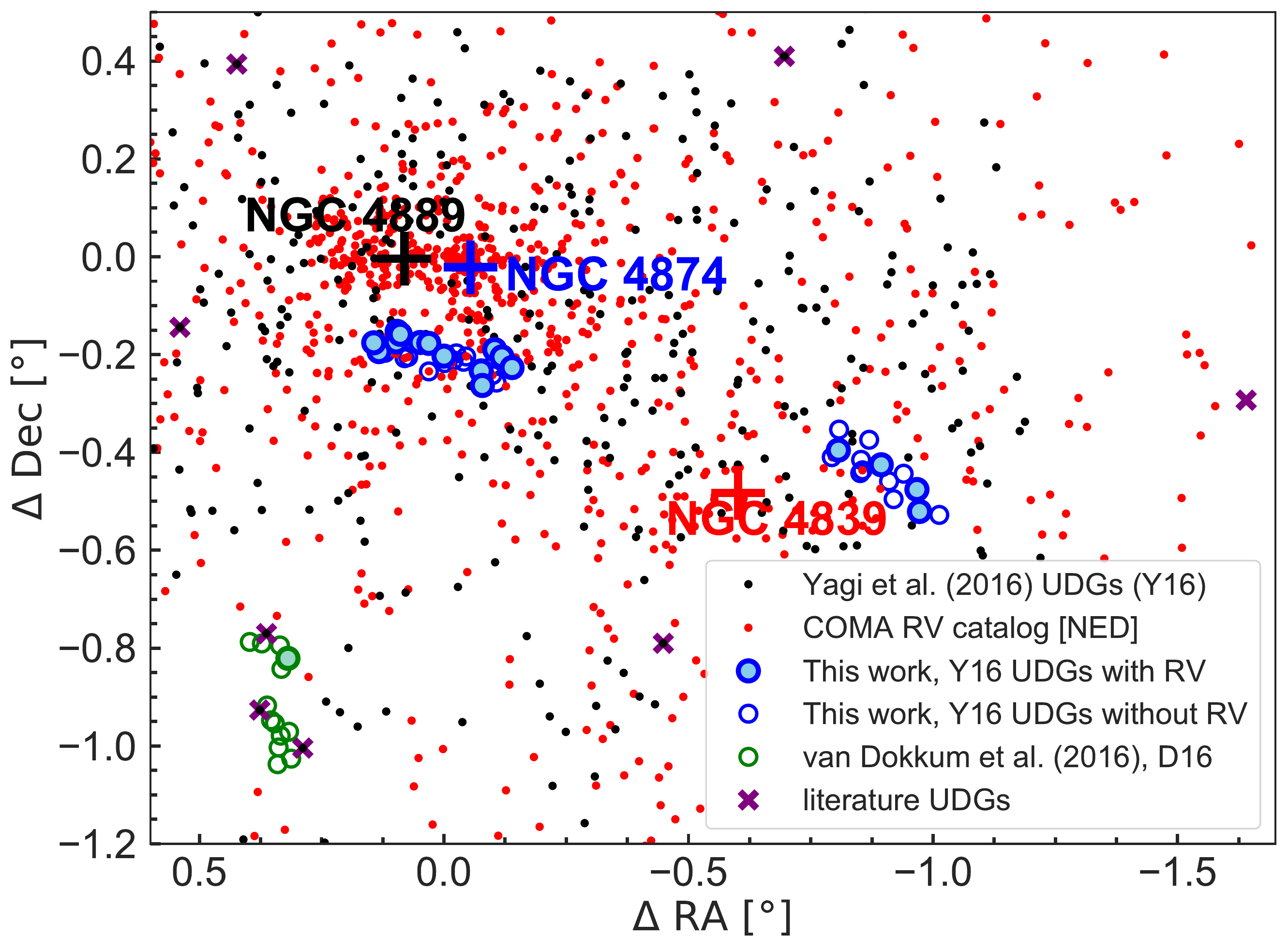}\hspace{0.01\textwidth}\\
	\caption{\label{fig:spatial_distr} Spatial distribution of galaxies in the Coma cluster, highlighting the central and outer fields studied in this work. The solid crosses show the 3 most luminous galaxies within the cluster, while the black dots are the UDGs from the Coma cluster LSB catalog of \citet{Yagi_2016}. The red dots are Coma cluster galaxies with radial velocity measurements available in the literature (sourced from NED). Spectroscopically confirmed Coma cluster UDGs from the literature are shown as purple crosses. The filled and open circles are the UDGs and non-UDGs, respectively, studied in this work for which we have successfully measured radial velocities. The green circles are from the DEIMOS observation (see Section \ref{D16} for more details) described in \citet{vanDokkum_2016}.}
\end{figure}

In Figure \ref{fig:sample_fig}, where we show the size--luminosity diagram of our target sample, most of our spectroscopic targets occupy a similar parameter space comparable to other Coma UDGs previously studied in the literature. 3 galaxies in our sample (Yagi266, Yagi413, and Yagi772) are not consistent with the generally accepted definition of UDGs. A careful comparison of the sizes of the UDGs common to both \hyperlink{Y16}{Y16} and the \citet{vanDokkum_2015} show that, on average, $\re$ from \hyperlink{Y16}{Y16} are ${\sim}15$~per cent smaller than those reported in \citet{vanDokkum_2015}, hence, we relax the UDG size criterion for the \hyperlink{Y16}{Y16} galaxies to $\re > 1.3$~kpc, and note that this size refers to the semimajor axis effective radius. This disparity in $\re$ could be due to differences in the image qualities or the galaxy size fitting techniques employed in both studies. Figure \ref{fig:spatial_distr} shows our spectroscopic targets in the plane of the sky. Unlike the spectroscopically confirmed Coma UDGs in the literature which are mostly at large projected clustercentric radii, our UDGs sample extends well into the cluster core.

\subsection{Observations}
\label{main_obs}

\begin{figure}
    \includegraphics[width=0.48\textwidth]{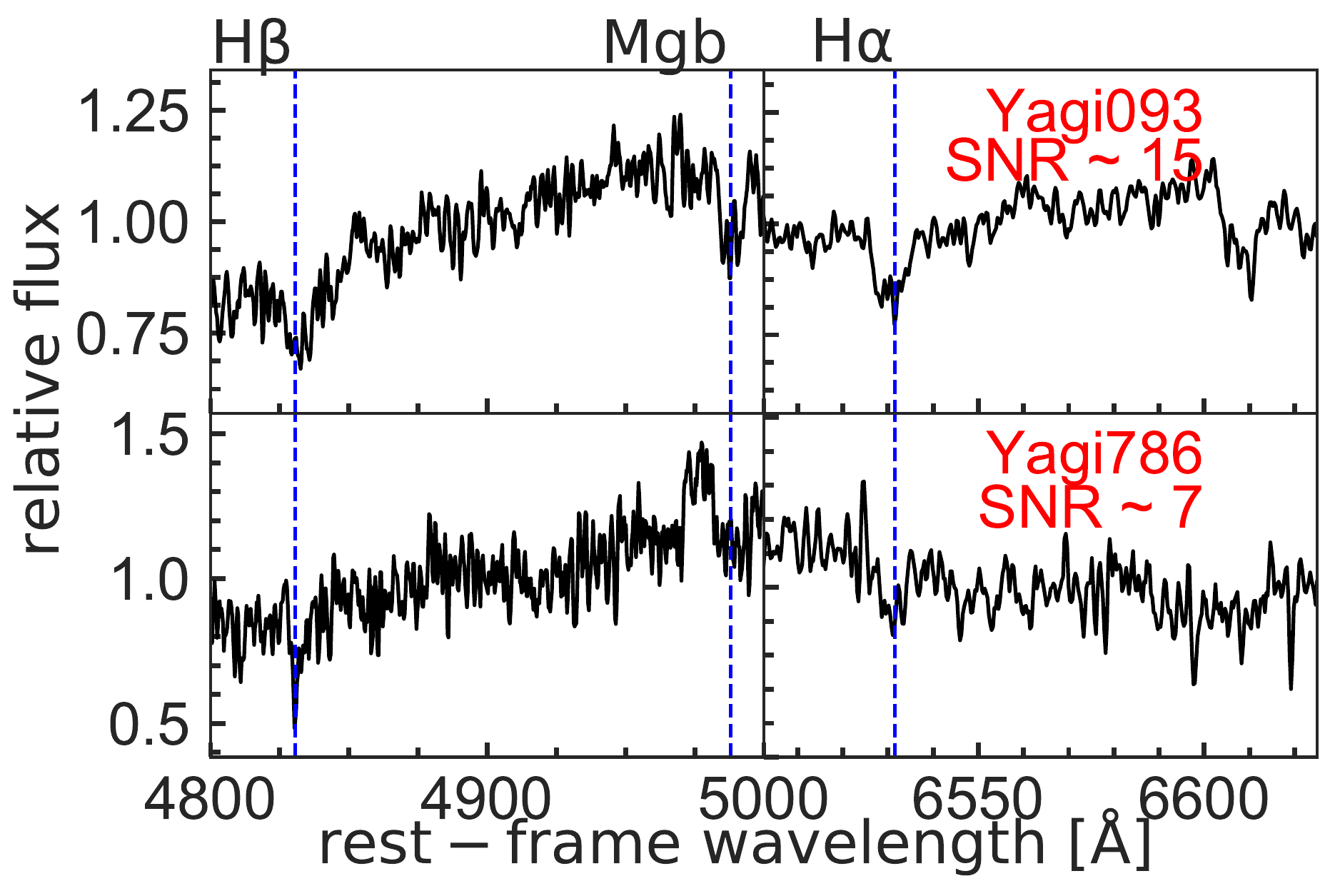}\hspace{0.01\textwidth}\\
	\caption{\label{fig:sample_spec} Rest-frame spectra of representative UDGs from the central and outer masks. In the \textit{top} and \textit{bottom} panels, we show the spectra of Yagi093 (central mask) and Yagi786 (outer mask), respectively, with the $\rm H{\beta}$, Mgb and $\rm H{\alpha}$ spectral absorption features highlighted.}
\end{figure}
Our spectroscopic data were obtained from the Keck II telescopes with the DEIMOS spectograph from 2017 April 27 - 29 during dark night conditions with a mean seeing of $0\farcs7$. We observed two DEIMOS masks, positioned at 0.4~Mpc in the south direction (central mask) and 1.6~Mpc in the south-west direction (outer mask), respectively, from the cluster center.
We set up DEIMOS with the GG455 filter and the 600~$\rm lines\ mm^{-1}$ grating centered on 6000~$\AA$. This resulted in spectral data with resolution of 14~$\AA$ (this corresponds to an instrumental resolution of ~300~$\rm km\ s^{-1}$) and wavelength coverage spanning ${\sim}4300 - 9600\ \AA$, depending on the position of the slit on the masks. We integrated on the central mask for a total of 14.5~hrs over 3~nights and on the outer mask for a modest total time of 2~hrs, with individual exposures of 30~minutes. With the longer integration on the central mask, we also explore the stellar populations of these UDGs in a companion paper (see Paper II).

Our masks had custom slits that were $3\arcsec$ wide in order to capture as much UDG light as possible. In addition, we also increased the gaps between slits from the nominal $0\farcs7$ to $1\arcsec$ and avoided placing bright filler objects on adjacent slits to our UDG targets to prevent possible cross-talk across slits. These adjustments are based on trial observations we made in January, 2017. To summarize our masks based on their targets, the central mask had 12 UDG slits, 2 LSB slits, 8 dedicated sky slits, and 15 filler-object slits, while the outer mask had 9 UDG slits, 1 LSB slit, 5 dedicated sky slits, and 23 filler-object slits. We identified suitable filler-objects from the field-of-view and placed slits on them so as not to waste any real-estate on the masks. A few of these filler-objects have been detected in the Sloan Digital Sky Survey (SDSS) and have kinematics data available in the literature with which we compare our new velocity measurements.

\subsection{Data Reduction}
\label{Data_redox}
We reduced the data with a modified version of the \textsc{idl DEEP2 DEIMOS spec2d} pipeline \citep{Cooper_2012} that accounts for the spatially diffuse nature of our targets and for proper sky subtraction. We performed two sets of data reduction on the central mask: one, where we combined the optimally-extracted 1D spectra data from each night, and two, where we combined the 2D slit spectra from the data reduction process before extracting a single final 1D spectra. We do not observe any significant difference between the final 1D spectra from the two methods, and hence, we use the reduced data from the first approach in subsequent analyses.

\begin{table*}
{\caption{Ultra-diffuse galaxy sample and other targets from the Coma cluster studied in this work. The horizontal lines separates UDGs from low surface brightness galaxies and from filler, non-UDGs, respectively. The galaxy IDs, coordinates, $R$-band magnitudes, $B-R$ colors, surface brightnesses and sizes for the UDGs and LSB galaxies are from the \citet{Yagi_2016} catalog. Mask ``C'' and ``O'' denote the central and outer masks, respectively, described in Section \ref{main_obs}, while mask ``D16'' is described in Section \ref{D16}. SNR is the signal-to-noise ratio measured around the $\rm H{\beta}$ or $\rm H{\alpha}$ region, as described in the text. The galaxy IDs shown for the filler objects are mostly from the SDSS DR14 \citep{SDSS} catalog except when they have also been cataloged in \citet{Godwin_1983}. We have translated their SDSS-$r$ band magnitudes into Subaru-$R$ band using eqn. $2$ and table~$1$ from \citet{Yagi_2013}. The sizes shown for the filler galaxies are from single component \Sersic/ fits to their Subaru-$R$ band image. Only the filler objects for which we have successfully measured radial velocities are shown in the table. In the `Add. IDs'' column, we list the IDs of our galaxy sample as presented in \citet{Godwin_1983} and \citet{vanDokkum_2015}.} \label{tab:data_summ}
\scalebox{0.92}{
\begin{tabular}{@{}l c c l l l l l l l l}
\hline
\hline
Galaxy & RA & Dec & Mask  & $R$     & $B-R$ & $\mu_{_{0}}$   & $\re$  & SNR & Vel & Add. IDs\\
 	   & [J2000] & [J2000] & 	   & [mag]  & [mag]  & [$\SBunit$]  & [kpc]	&	 &[$\kms$] & \\
\hline
  Yagi090 & 13:00:20.37 & +27:49:24.0 & C & 20.3 & 0.86 & 24.4 & 1.92 & $<5$ & $-$ &\\ 
  Yagi093 & 13:00:20.61 & +27:47:12.3 & C & 18.9 & 0.96 & 23.6 & 3.49 & 15.3 & 6611$\pm$\,137 & DF26, GMP2748\\
  Yagi098 & 13:00:23.20 & +27:48:17.1 & C & 19.6 & 0.96 & 24.9 & 2.30 & 20.8 & 5980$\pm$\,82 &\\
  Yagi263 & 12:59:15.33 & +27:45:14.8 & C & 20.8 & 1.04 & 24.3 & 2.04 & 5.7 & 6695$\pm$\,147 &\\
  Yagi275 & 12:59:29.89 & +27:43:03.1  & C & 19.2 & 0.92 & 23.5 & 2.93 & 15.2 & 4847$\pm$\,149 & GMP3418\\ 
  Yagi276 & 12:59:30.46 & +27:44:50.4 & C & 19.6 & 0.9  & 24.0 & 2.25 & 13.9 & 7343$\pm$\,102 & DF28\\
  Yagi285 & 12:59:48.72 & +27:46:39.0 & C & 19.7 & 1.0  & 24.8 & 3.88 & 12.4 & 6959$\pm$\,121 & DF25\\ 
  Yagi364 & 12:59:23.85 & +27:47:27.2 & C & 20.0 & 0.95 & 25.2 & 2.42 & 15.8 & 7068$\pm$\,90 & DF23\\ 
  Yagi392 & 12:59:56.17 & +27:48:12.8 & C & 20.7 & 0.97 & 24.0 & 1.46 & 7.3 & 7748$\pm$\,161 & \\ 
  Yagi398 & 13:00:00.41  & +27:48:19.7 & C & 20.1 & 0.96 & 23.6 & 1.34 & 19.5 & 4180$\pm$\,167 & \\ 
  Yagi417 & 13:00:12.10 & +27:48:23.5 & C & 21.4 & 0.83 & 24.9 & 1.62 & 8.5 & 9038$\pm$\,179 & \\ 
  Yagi418 & 13:00:11.71 & +27:49:41.0 & C & 20.4 & 0.93 & 23.9 & 1.57 & 16.0 & 8335$\pm$\,187 & \\ 
  Yagi762 & 12:55:55.40 & +27:27:35.9 & O & 20.5 & $-$  & 24.5 & 1.97 & 5.5 & 7188$\pm$\,127 & DF36\\ 
  Yagi764 & 12:55:56.65 & +27:30:17.6 & O & 20.0 & $-$  & 23.5 & 2.44 & 7.2 & 7050$\pm$\,115 & \\
  Yagi767 & 12:55:59.15 & +27:25:53.4 & O & 20.6 & $-$  & 24.3 & 2.37 & $<5$ & $-$ & \\
  Yagi771$^a$ & 12:56:05.38  & +27:30:18.2 & O & 22.2 & $-$ & 24.8 & 1.60 & 5.4 & 11007$\pm$\,192 & \\
  Yagi774 & 12:56:12.95 & +27:32:50.3 & O & 20.6 & 0.68 & 25.9 & 3.07 & $<5$ & $-$ & \\ 
  Yagi776 & 12:56:14.15 & +27:33:19.7 & O & 20.2 & 0.81 & 24.0 & 2.3  & 7.4 & 8473$\pm$\,81 & \\ 
  Yagi781 & 12:56:28.29 & +27:36:11.5 & O & 20.8 & $-$  & 24.2 & 1.64 & $<5$ & $-$ & \\
  Yagi782 & 12:56:28.41 & +27:37:06.3  & O & 20.1 & $-$  & 24.6 & 2.8  & $<5$ & $-$ & DF32\\ 
  Yagi786 & 12:56:35.20 & +27:35:07.0  & O & 19.4 & $-$  & 23.5 & 2.34 & 6.6 & 7810$\pm$\,141 & \\
  Yagi012$^b$& 13:01:05.30 & +27:09:35.1 & D16 & 20.6 & $-$  & 23.7 & 1.57 & 4.1 & 6473$\pm$\,33 & DFX2 \\ 
\hline
  Yagi266 & 12:59:20.25 & +27:46:33.3 & C & 22.6 & $-$  & 25.7 & 0.87 & 5.2 & 6366$\pm$\,139 &\\
  Yagi413 & 13:00:10.19 & +27:49:19.8 & C & 22.2 & 0.84 & 24.2 & 0.85 & 9.7 & 5014$\pm$\,196 & \\ 
  Yagi772 & 12:56:09.07  & +27:34:16.2 & O & 21.7 & $-$  & 24.7 & 1.21 & $<5$ & $-$ & \\
\hline
  GMP2749 & 13:00:20.48 & +27:48:17.0 & C & 18.4 & $-$  & $-$ & 1.59 & 18.8 & 5846$\pm$\,74 & \\ 
  GMP2800 & 13:00:17.55 & +27:47:03.9  & C & 16.7 & $-$  & $-$ & 2.92 & 21.8 & 7001$\pm$\,132 & \\
  GMP2923 & 13:00:08.05  & +27:46:24.1 & C & 16.8 & $-$  & $-$ & 2.09 & 9.8 & 8652$\pm$\,125 & \\
  GMP2945 & 13:00:06.29  & +27:46:32.9 & C & 14.6 & $-$  & $-$ & 2.63 & 24.3 & 6091$\pm$\,66 & \\ 
  GMP3037 & 12:59:59.39 & +27:47:55.8 & C & 19.1 & $-$  & $-$ & 0.76 & 11.6 & 13938$\pm$\,142 & \\ 
  GMP3071 & 12:59:56.11 & +27:44:46.7 & C & 16.1 & $-$  & $-$ & 1.39 & 13.9 & 8810$\pm$\,99 & \\
  GMP3519 & 12:59:22.94 & +27:43:24.5 & C & 18.7 & $-$  & $-$ & 1.88 & 19.2 & 4062$\pm$\,167 & \\ 
  GMP3298 & 12:59:37.83 & +27:46:36.6 & C & 15.3 & $-$  & $-$ & 4.27 & 18.5 & 5554$\pm$\,41 & \\ 
  GMP3493 & 12:59:24.93 & +27:44:19.9 & C & 14.9 & $-$  & $-$ & 1.38 & 23.2 & 6001$\pm$\,80 & \\ 
  J125924.95+274529.0$^c$ & 12:59:24.95 & +27:45:29.0 & C & 21.9 & $-$  & $-$ & 0.82 & 6.2 & 5420$\pm$\,166 & \\ 
  J125944.10+274607.5 & 12:59:44.11 & +27:46:07.6  & C & 19.2 & $-$  & $-$ & 0.91 & 22.9 & 6109$\pm$\,127 & \\ 
  J125942.65+274658.8 & 12:59:42.65 & +27:46:59.4 & C & 20.2 & $-$  & $-$ & 1.36 & 30.1 & 5418$\pm$\,163 & \\
  J125948.33+274547.6 & 12:59:48.37 & +27:45:48.2 & C & 21.2 & $-$  & $-$ & 1.37 & 9.6 & 8039$\pm$\,109 & \\
  J125939.09+274557.5 & 12:59:39.10 & +27:45:57.5 & C & 20.1 & $-$  & $-$ & 0.88 & 15.7 & 7791$\pm$\,164 & \\
  GMP5357 & 12:56:34.86 & +27:37:38.6 & O & 18.5  & $-$ & $-$ & 1.98 & 13.2 & 7239$\pm$\,97 & \\
  GMP5455 & 12:56:24.46 & +27:32:18.4 & O & 17.8  & $-$ & $-$ & 1.33 & 18.4 & 7413$\pm$\,85 & \\
  GMP5465 & 12:56:23.36 & +27:32:38.5 & O & 17.3  & $-$ & $-$ & 0.87 & 17.5 & 7149$\pm$\,50 & \\
  J125638.44+273415.3 & 12:56:38.44 & +27:34:15.3 & O   & 17.4 & $-$  & $-$ & 1.68 & 8.8 & 7549$\pm$\,49 & \\
  J125623.18+273358.1 & 12:56:23.18 & +27:33:58.1 & O   & 19.2 & $-$  & $-$ & 0.68 & 5.0 & 5655$\pm$\,69 & \\  
  J125608.14+272906.5 & 12:56:08.14 & +27:29:06.5 & O   & 21.7 & $-$  & $-$ & 1.07 & 5.1 & 6946$\pm$\,143 & \\
  J125620.09+273623.7 & 12:56:20.09 & +27:36:23.7 & O   & 20.7 & $-$  & $-$ & 0.97 & 13.3 & 6208$\pm$\,87 & \\
  J125610.30+273104.4 & 12:56:10.30 & +27:31:04.4 & O   & 22.7 & $-$  & $-$ & 0.44 & 5.2 & 6333$\pm$\,182 & \\
  J125603.20+273213.4 & 12:56:03.20 & +27:32:13.4  & O   & 21.4 & $-$  & $-$ & 0.57 & 8.7 & 3434$\pm$\,150 & \\
  J130110.36+265636.4 & 13:01:10.36 & +26:56:36.4 & D16 & 18.3 & $-$  &$-$  & 0.53 & 10.6 & 7916$\pm$\,75 & \\
  J130103.69+265717.2 & 13:01:03.69 & +26:57:17.2 & D16 & 19.3 & $-$  &$-$  & 0.38 & 14.6 & 7719$\pm$\,40 & \\
  J130109.88+265839.2 & 13:01:09.88 & +26:58:39.2  & D16 & 21.3 & $-$  &$-$  & 0.49 & 6.2 & 9028$\pm$\,25 & \\
  J130108.86+270006.3 & 13:01:08.86 & +27:00:06.3  & D16 & 22.7 & $-$  &$-$  & 0.40 & 4.8 & 5272$\pm$\,40 & \\
  J130104.75+270035.0 & 13:01:04.75 & +27:00:35.0 & D16 & 18.1 & $-$  &$-$  & 0.71 & 16.2 & 4563$\pm$\,67 & \\
  J130111.71+270136.9 & 13:01:11.71 & +27:01:36.9 & D16 & 21.2 & $-$  &$-$  & 0.65 & 7.7 & 4160$\pm$\,27 & \\
  J130113.58+270158.8 & 13:01:13.58 & +27:01:58.8 & D16 & 22.0 & $-$  &$-$  & 0.4  & 8.8 & 8015$\pm$\,37 & \\
  J130115.54+270348.0 & 13:01:15.54 & +27:03:48.0 & D16 & 19.9 & $-$  &$-$  & 0.47 & 12.6 & 7762$\pm$\,58 & \\
  J130108.49+270817.7 & 13:01:08.49 & +27:08:17.7 & D16 & 20.7 & $-$  &$-$  & 0.86 & 7.7 & 5045$\pm$\,34 & \\
  J130109.18+271110.3 & 13:01:09.18 & +27:11:10.3 & D16 & 21.0 & $-$  &$-$  & 0.49 & 6.3 & 9957$\pm$\,29 & \\
  J130117.24+271104.5 & 13:01:17.24 & +27:11:04.5 & D16 & 21.4 & $-$  &$-$  & 0.34 & 4.2 & 5058$\pm$\,20 & \\
  J130124.47+271129.5 & 13:01:24.47 & +27:11:29.5 & D16 & 22.5 & $-$  &$-$  & 0.47 & 6.6 & 8428$\pm$\,44 & \\ 
\hline
\end{tabular}}
\scriptsize
\begin{flushleft}
{\textit{a}. The size we report for Yagi~771 is updated to correspond to a distance of ${\sim}160$~Mpc and the radial velocity shown.\\
 \textit{b}. Yagi012 is the ultra-diffuse galaxy from the D16 mask with radial velocity not published in the literature.\\
 \textit{c}. J125924.95+274529.0 is not detected in SDSS DR14 but has photometry available in the catalog of \citet{Adami_2006}.}
\end{flushleft}}
\end{table*}

\begin{figure*}
    \includegraphics[scale=0.34]{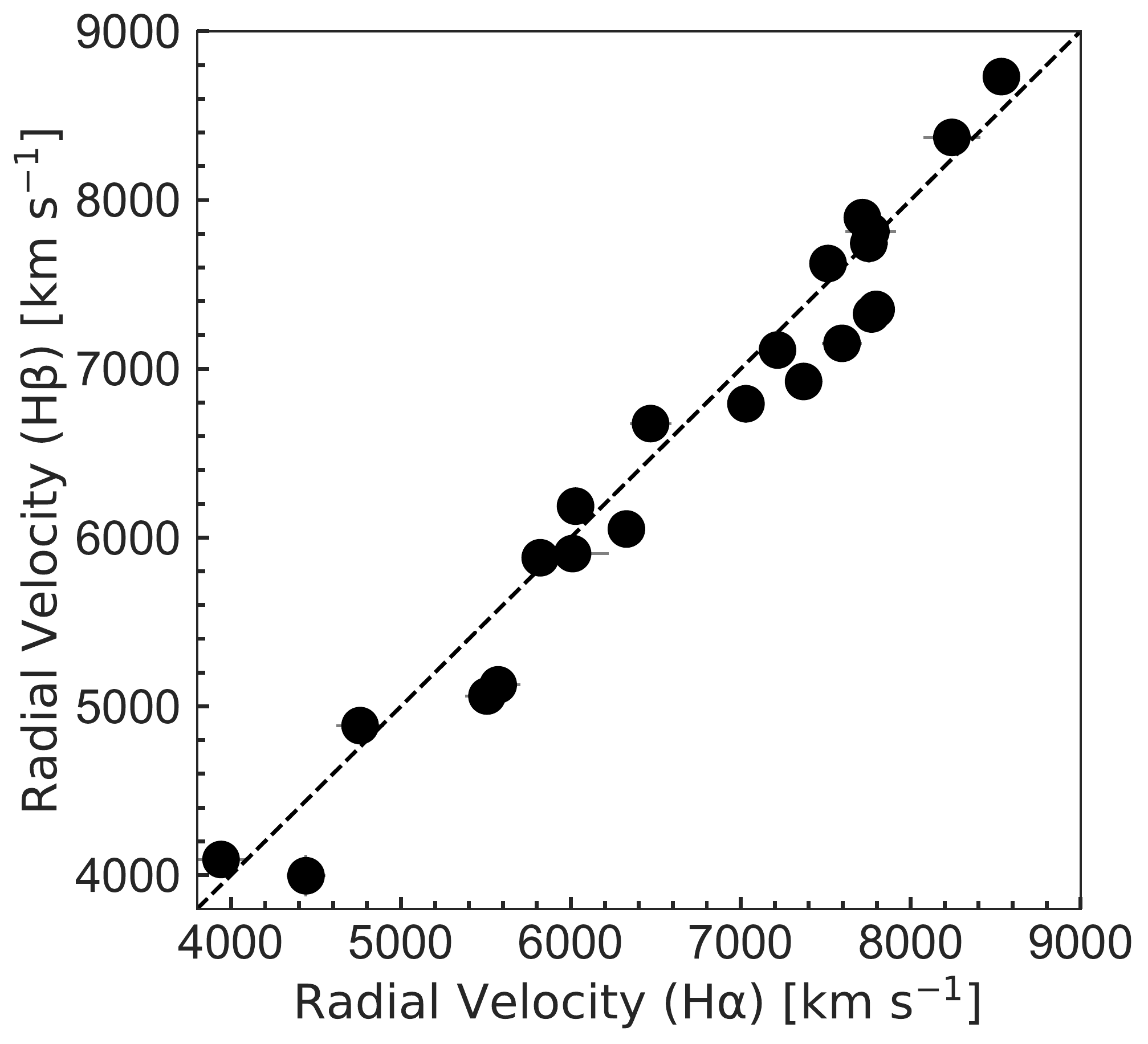}\hspace{0.01\textwidth}%
    \includegraphics[scale=0.34]{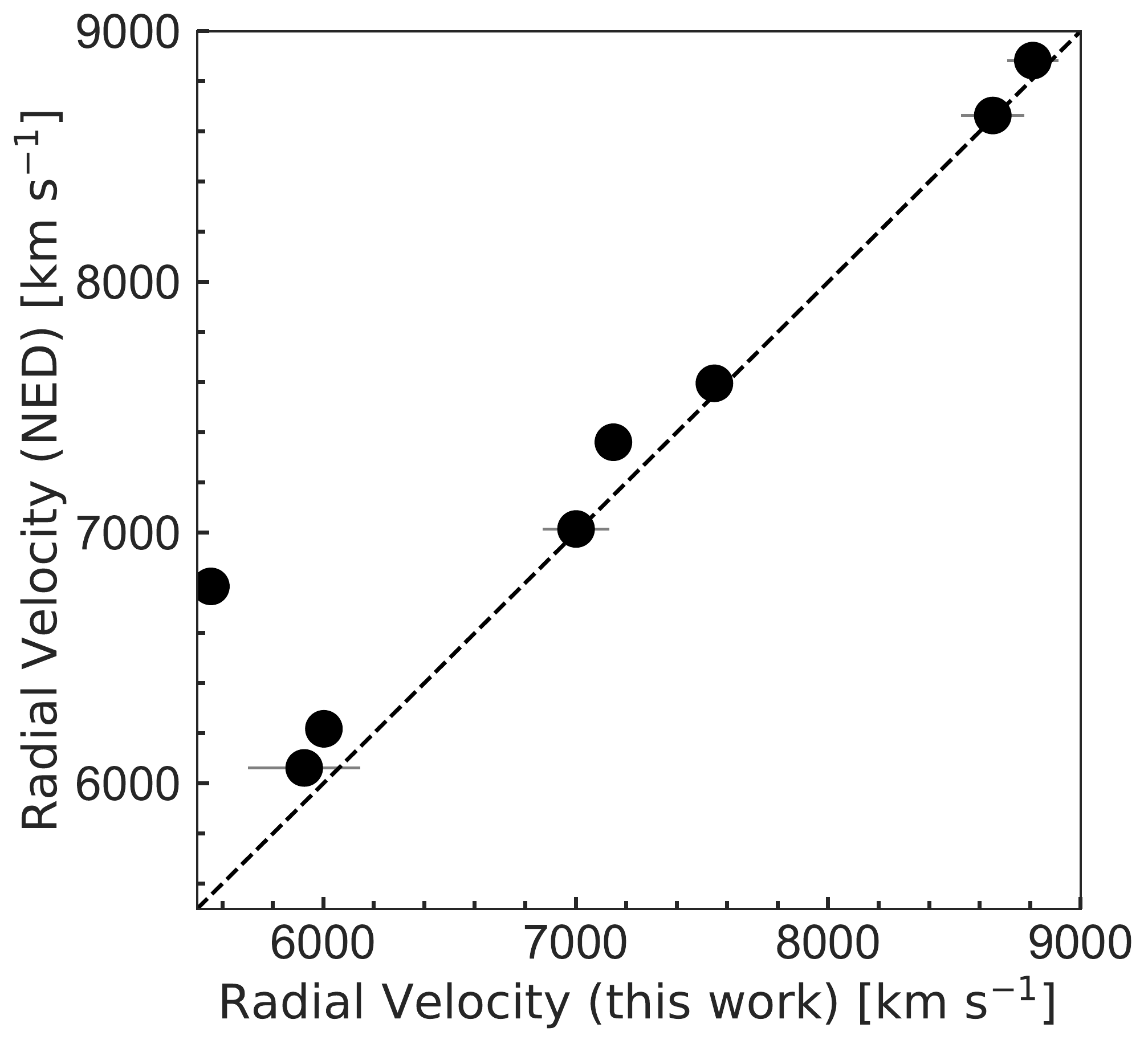}\hspace{0.01\textwidth}
	\caption{\label{fig:cmp_Res} \textit{Left}: Comparison of radial velocities obtained around the $\rm H{\beta}$ and $\rm H{\alpha}$ spectroscopic features. \textit{Right}: Comparison of radial velocities obtained in this work with measurements available in the literature. The significantly deviant measurement is due to bad wavelength calibration in our spectral data for GMP3298, where only the H$\alpha$ spectral range is available.}
	\vspace*{\floatsep}
	\includegraphics[width=0.94\textwidth]{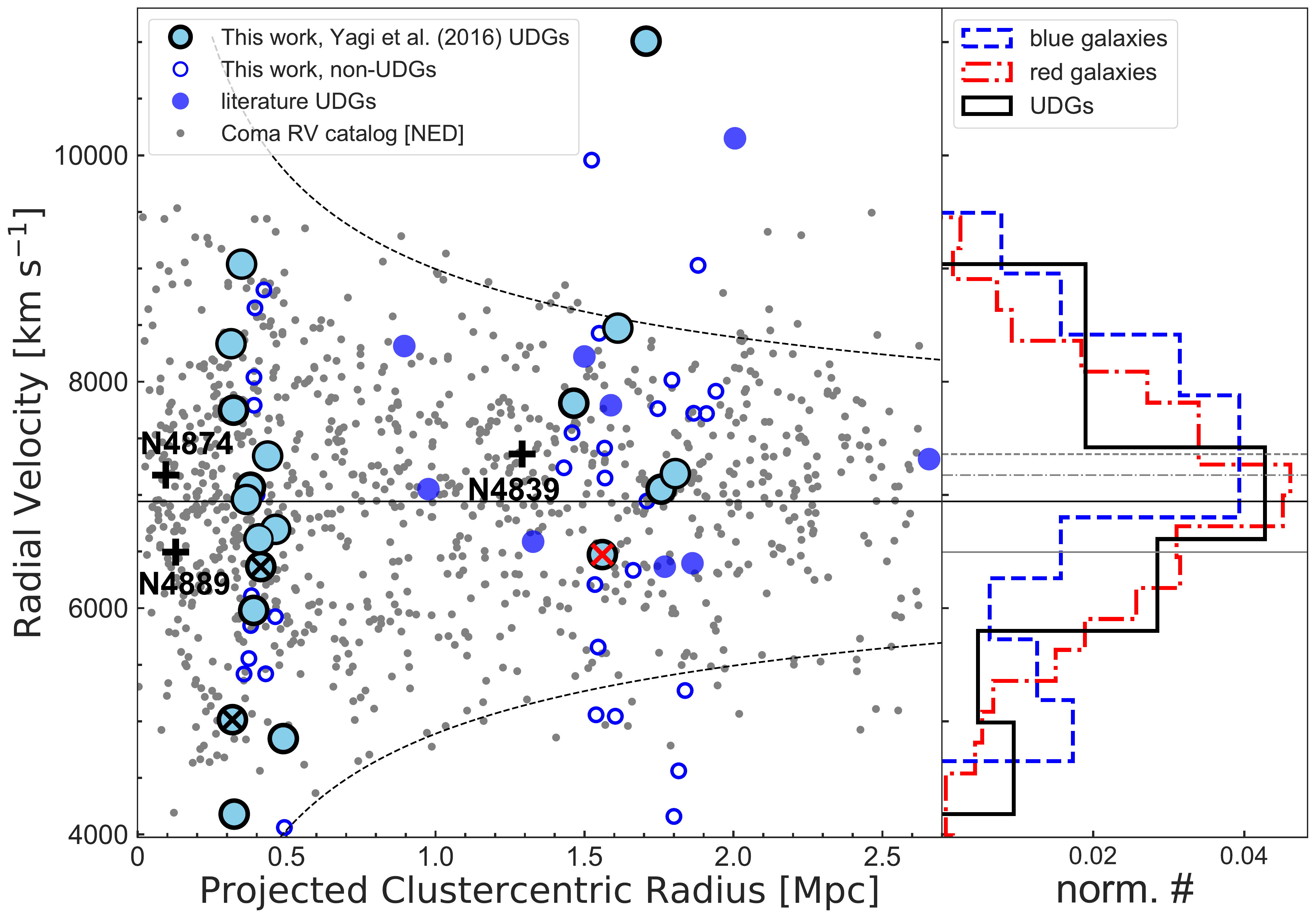}\hspace{0.01\textwidth}\\
	\caption{\label{fig:RV_distr} Clustercentric radial velocity distribution of galaxies in the Coma cluster. In the \textit{left} panel, cluster galaxies with radial velocity measurements available in the literature are shown as gray dots. The black solid line is the recessional velocity of the Coma cluster, while the dashed curves shows the velocity limits consistent with the virial mass reported in \citet{Kubo_2007}. We have marked out the 3 most luminous galaxies in the Coma cluster, i.e., NGC~4874, NGC~4889 and NGC~4839. UDGs from \citet{Yagi_2016} for which we have measured radial velocities are shown as the skyblue circles with black edges. The blue open circles are the non-UDGs for which we have also obtained radial velocities. The blue circles without black edges are UDGs with radial velocity measurements from the literature. Yagi012, the UDG from the D16 mask, is marked with a red cross, while the low surface brightness galaxies (see Table \ref{tab:data_summ}) are marked with black crosses. Yagi~771, one of our target UDGs, is consistent with being a cluster background galaxy. The \textit{right} panel show normalized velocity histograms of the red ($B-R > 0.87$), blue ($B-R \leq 0.87$) cluster galaxy subgroups compared to our UDG sample. The gray lines show the velocity of the 3 most luminous galaxies in the Coma cluster.}
\end{figure*}

\section{Analysis and Results}
\label{analysis}
\subsection{Radial Velocities}
\label{rad_vel}
The reduced 1D-spectra from the central and outer masks have signal-to-noise ratio (SNR) ranging over 5--20 and 5--7~${\AA}^{-1}$, respectively. Owing to this modest SNR, we use the \textsc{fxcor} package in \textsc{iraf} to determine the radial velocities of our targets, though we also used the \textsc{reduceme/midez} package \citep{Cardiel_1999} on the central mask to independently determine our radial velocities. Both methods returned radial velocities that are in excellent agreement. In \textsc{iraf}, we cross-correlated our 1D-spectra with stellar templates sourced from the new MILES \citep{Vazdekis_2016} spectral library (with FWHM of 14~\AA), obtained the cross-correlation peak, and fitted it with a sinc function\footnote{We also tried the cross-correlation using the parabolic function but we only report results that are invariant to the fitted function.} to obtain the velocity estimates.

To obtain robust velocity estimates, we only consider outputs from \textsc{fxcor} with the Tonry \& Davis ratio (TDR parameter; \citealt{Tonry_1979}) $\geq 3$. This TDR parameter cut-off is equivalent to a SNR~$\geq5$ limit. Furthermore, when both the $\rm H{\beta}$ and $\rm H{\alpha}$ absorption features are observed in the spectral data, as is the case for 23 out of the 62 science slits, we split the spectrum into two segments, i.e., ${\sim}4600 - 5100$ and ${\sim}6400 - 6900$~$\AA$, respectively, and obtained independent estimates of the redshifts. We note that due to the relative brightness of the night sky, we are not able to use our reduced spectral data redward of 6900 \AA. Figure \ref{fig:sample_spec} shows the rest-frame spectra of two representative UDGs from both masks.

The offsets between these multiple velocity estimates vary between $10$ and $450 {\rm km\ s^{-1}}$. We note that this is adequate for the purpose of determining the cluster membership of our targets, and consistent with our spectral resolution. In Figure \ref{fig:cmp_Res}, we show the agreement between the velocity estimates obtained from the $\rm H{\beta}$ and $\rm H{\alpha}$ spectral regions. For these targets, we adopt the weighted average of the $\rm H{\beta}$ and $\rm H{\alpha}$-determined velocities, weighting by the TDR parameter returned from \textsc{fxcor}. Otherwise, we report velocity measurements made around the $\rm H{\beta}$ region. We obtained uncertainties on our velocity measurements by summing in quadrature the \textsc{fxcor} error and the standard deviation among the stellar templates. We are able to compare the radial velocities of some filler objects with measurements available in the literature and show the generally good agreement in Figure \ref{fig:cmp_Res}.

\subsection{Remarks on the D16 mask}
\label{D16}
As shown in Figure \ref{fig:spatial_distr}, we have also supplemented our data with the non-UDG objects from the DEIMOS mask (D16) reported in \citet{vanDokkum_2016, vanDokkum_2017}. This mask was observed with the GG550 filter, the 1200~$\rm lines\ mm^{-1}$ grating and a central wavelength of 6300~$\AA$ for a total of 33.5~hrs over two nights. The mask was designed primarily to observe 4 UDGs: two from the \citet{vanDokkum_2015} catalog, i.e., DF42 and DF44, and two additional UDGs from their CFHT imaging introduced in \citet{vanDokkum_2017}, i.e., DFX1 and DFX2. Note that DFX2 is also cataloged in \hyperlink{Y16}{Y16} as Yagi012. Apart from Yagi012, these UDGs all have their radial velocities published in \citet{vanDokkum_2017}. In addition, 33 slits were positioned on mask filler objects.

We reduced the entire mask as described in Section \ref{Data_redox} except in this case, we perform local sky-subtraction. While such sky-subtraction might not be adequate for the UDGs on this mask due to their diffuse nature, however, the sole purpose of adding this mask is to obtain radial velocities for the non-UDG cluster galaxies near the UDGs. We successfully measured the radial velocities of 25 (including all the UDGs) out of the 33 slit targets as described in Section \ref{rad_vel}, using only the $\rm H{\alpha}$ spectral region. We measured a radial velocity of $6473$~$\kms$ for Yagi012, and similar velocities for the other UDGs as already published in the literature. The non-UDGs with radial velocities consistent with the Coma cluster are shown in Table \ref{tab:data_summ}.

\subsection{Summary of results}
We summarize our heliocentric-corrected radial velocity measurements, and other relevant data, in Table \ref{tab:data_summ} and show the clustercentric radial velocity distribution in Figure \ref{fig:RV_distr}. From this diagram, a few spectroscopically confirmed cluster galaxies in the cluster outskirts (both from our new measurements and our NASA/IPAC
Extragalactic Database (NED)\footnote{http://ned.ipac.caltech.edu/} compilation) have radial velocities that often deviate from the cluster velocity limits shown. These may be new infalls or back-splashing galaxies. We have obtained reliable radial velocity measurements for 19 out of our original sample of 25 LSB galaxies, with 16 of them being UDGs and 2 LSB galaxies. One of the UDGs in the outer mask, Yagi~771, has a velocity consistent with being a cluster background galaxy. Our velocity measurement puts it at ${\sim}60$~Mpc behind the Coma cluster. This adds to the list of field UDGs with spectroscopic velocity measurements, the others being DGSAT~I \citep{Martinez_2016} and, DF03 \citep{Kadowaki_2017}. We have updated its $\re$ in Table \ref{tab:data_summ} to reflect this new result. In addition, we confirm the cluster membership of Yagi012, from the D16 mask we have reduced. We have also obtained radial velocities for 35 non-UDG Coma cluster galaxies. Also, there are 3 UDGs (DF42, DF44, and Yagi012) in the South-East direction of the cluster (see Figure \ref{fig:spatial_distr}) that are within 320~kpc and 130~\kms of each other, and thus may represent abound group of UDGs.

\section{Discussion}
\label{discussion}
In this work, we have successfully measured the radial velocities of 16 ultra-diffuse galaxies, 2 low surface brightness galaxies, and one background cluster galaxy from the \hyperlink{Y16}{Y16} Subaru low surface brightness catalog. This brings the total number of Coma cluster UDGs with radial velocity measurements to 24 (see \citealt{Kadowaki_2017, Meng_2017, vanDokkum_2017}), with the important addition that we now report spectroscopically confirmed UDGs within the projected cluster core, i.e. $\leq 0.5$~Mpc. With our UDG kinematic data that span a wide clustercentric radial baseline, we now briefly address the following salient questions that should help provide more insights into the origins of UDGs in cluster environments.

\subsection{Are ultra-diffuse galaxies kinematically distinct within the cluster?}
As shown in Figure \ref{fig:RV_distr}, UDGs within the projected cluster core have a similar velocity range to other neighboring galaxies along the line-of-sight. The UDGs with highest relative line-of-sight velocities within the core are linked with the deepest parts of the cluster gravitational potential well. The situation is a little nuanced for the UDGs at larger clustercentric radii since we do not have complete azimuthal coverage. The kinematics of the UDGs from our outer mask most likely reflect the peculiar local effect within the substructure centered on NGC~4839. This well-known substructure has been identified in several spatio-velocity, X-ray profile, and stellar population studies of the Coma cluster (\citealt{Biviano_1996, Colless_1996, Briel_2001, Neumann_2003, Adami_2005, Smith_2009, Smith_2012}, etc.). The 4 UDGs from our outer mask have a mean radial velocity of ${\sim}7630\ \kms$, comparable with the recession velocity of NGC~4839, i.e., ${\sim}7360\ \kms$.  Likewise, the 4 UDGs from the D16 mask have a mean radial velocity of ${\sim}6835\ \kms$, similar to the mean velocity of their neighboring galaxies. Within the cluster core, the mean radial velocity of the UDGs is ${\sim}6800\ \kms$, comparable to the systemic velocity of the cluster ($6943\ \kms$), while they have a velocity dispersion of ${\sim}1350\ \kms$. NED cluster galaxies within the cluster core, on the other hand, have a mean radial velocity of ${\sim}6910\ \kms$ and a velocity dispersion of ${\sim}1110\ \kms$. Outside the cluster core, UDGs (including the literature UDGs) have a significantly elevated mean velocity of ${\sim}7302\ \kms$. Note that these results for our UDGs do not change siginficantly if we include the two LSB galaxies in our analysis.

From the histograms in Figure \ref{fig:RV_distr}, the virialization state of both the blue galaxies and our UDG sample are comparable (with mean velocities ${\sim}7204$ and ${\sim}7200\ \kms$, and velocity dispersions ${\sim}1145$ and ${\sim}1220\ \kms$, respectively), unlike the more relaxed red galaxy subpopulation (with mean velocities ${\sim}6930\ \kms$, and velocity dispersion ${\sim}950\ \kms$). Taking all these results together, and bearing in mind our limited azimuthal coverage, it appears that Coma UDGs may not be as relaxed as the red galaxy subpopulation within the cluster, and have kinematics consistent with their local neighborhood. These suggest a formation origin that is closely linked with the gravitational potential within their local environment. We explore these in more detail below.

\subsection{Are ultra-diffuse galaxies kinematically distinct? additional hints from optical colors and sizes}
To further investigate the impact of the local environment on present-day UDGs within the cluster environment, we expand our parameter space, exploring their optical colors as a function of clustercentric radii. We also compare our UDGs with a sample of co-spatial, well studied relatively high surface brightness (HSB) dwarf galaxies from \citet{Smith_2009}. Optical $B-R$ colors are directly available for 12 out of the 24 UDGs as published in \hyperlink{Y16}{Y16}, which they obtained from \citet{Yamanoi_2012}. Also, a few of the remaining spectroscopically confirmed UDGs, especially at large radii, have stellar population parameters available in the literature, from which we can infer their $B-R$ color. We have also done a match of the Coma galaxies with radial velocities available in NED with the SDSS DR14 catalog in order to obtain their SDSS $g-r$ optical colors. All the HSB dwarf galaxies from \citet{Smith_2009} have SDSS $g-r$ colors. To convert the SDSS~$g-r$ colors into Subaru filter system $B-R$, we extracted transformation equations from Fig. 3 in \citet{Yamanoi_2012} based on stars with color measurements common to both studies. The transformation equation we use is $B-R = 1.4\ (g-r)+0.05$. 

\begin{figure}
    \includegraphics[width=0.48\textwidth]{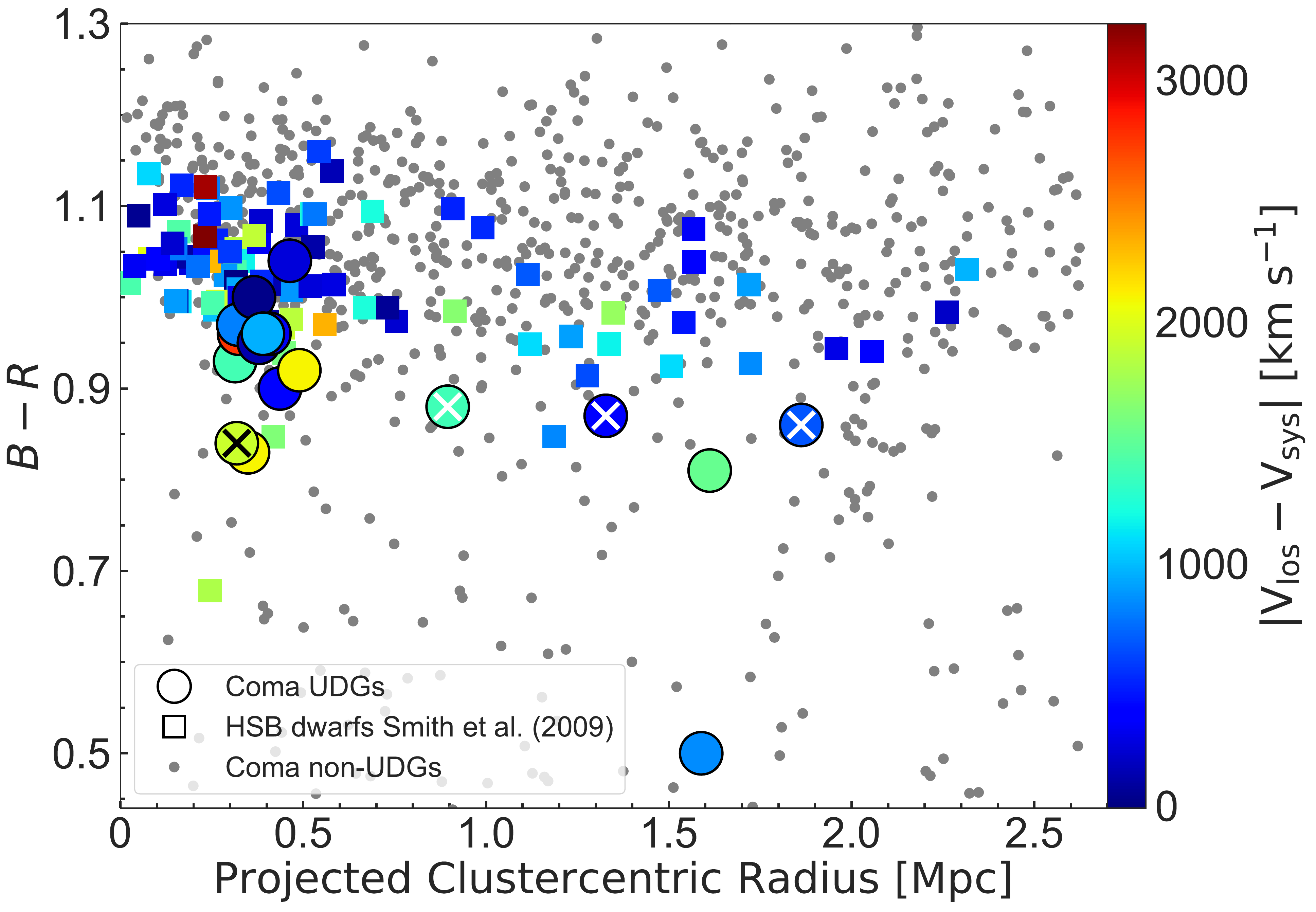}\hspace{0.01\textwidth}\\
	\caption{\label{fig:col_distr} Optical color as a function of clustercentric radii for spectroscopically confirmed galaxies in the Coma cluster. Ultra-diffuse galaxies without color information from \citet{Yamanoi_2012} and/or stellar population parameters in the literature have been left out of the plot. UDGs with inferred colors have been marked with white crosses. Galaxy types are as shown in the plot legend, with the ultra-diffuse galaxies, low surface brightness galaxies (marked with black cross), and the \citet{Smith_2009} dwarf galaxies color-coded by their absolute line-of-sight velocities relative to the cluster. Within the cluster core, UDGs are significantly less red compared to the co-spatial high surface brightness dwarfs, while at any clustercentric radius, galaxies with higher absolute relative velocity are bluer.}
\end{figure}
\begin{figure}
    \includegraphics[width=0.48\textwidth]{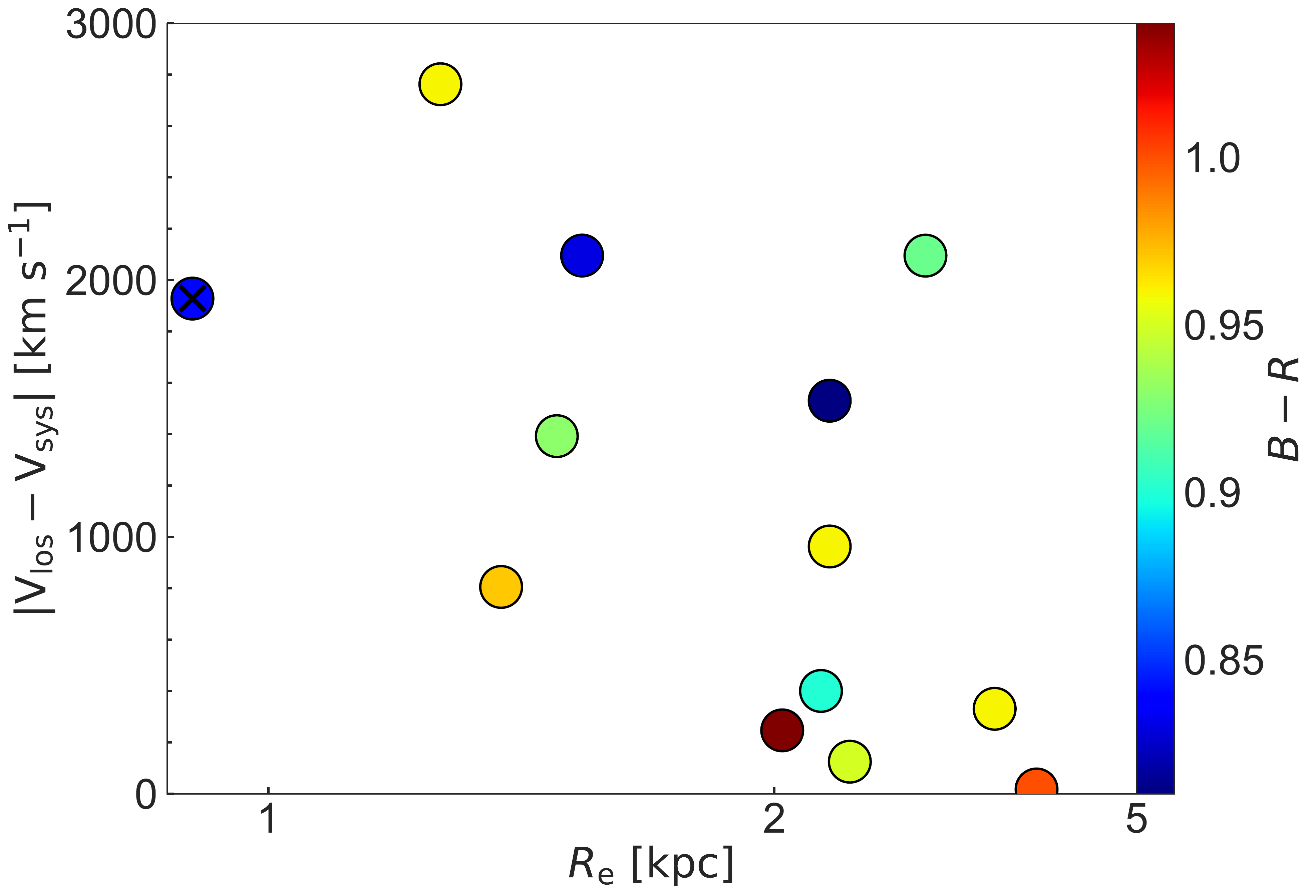}\hspace{0.01\textwidth}\\
	\caption{\label{fig:size_vel} Absolute relative radial velocities of Coma ultra-diffuse galaxies as a function of effective radii. We have color-coded the UDGs according to their optical $B-R$ colors and also added the low surface brightness galaxy (marked with black cross) with color information to this plot. Bigger and slower moving UGDs are seen to be redder.}
\end{figure}

From Figure \ref{fig:col_distr}, at the same clustercentric radii within the cluster core, slower moving UDGs and HSB dwarfs, relative to the cluster, are significantly redder in optical colors. Outside the cluster core, we only have two confirmed UDGs with observed colors, although we also show the inferred $B-R$ colors of the UDGs from \citet{Meng_2017}. The very blue UDG, with $B-R$ color of 0.5 is DF40 (also Yagi507 in \hyperlink{Y16}{Y16}) from \citet{Kadowaki_2017}, part of their spectral stack which they judged to be very old and metal-poor. Due to the paucity of our data here and the peculiar effects from the NGC~4839 substructure, the true nature of the color trend in the cluster outskirts is still vague. At the least, UDGs and HSB dwarfs also follow a velocity sequence, especially within the cluster core. Leaving the 3 UDGs from \citealt{Meng_2017} (they have a mean inferred $B-R$ color of $0.87$) out of Figure \ref{fig:col_distr} does not change this conclusion. In Figure \ref{fig:size_vel} where we show the absolute relative radial velocities of our UDGs (including the LSB galaxy) as a function of their semimajor--axis effective radii, we find tentative evidence that not only are the slower UDGs redder, but they are also bigger (and \textit{vice versa}). This might be the effect of tidal heating \citep{Boselli_2006} which makes them bigger assuming they were not formed in-situ already bigger. It is important to note that the results presented in this section are valid regardless of the LSB galaxy with color information.

\subsection{Are ultra-diffuse galaxies recent infalls into the cluster?}
\begin{figure}
    \includegraphics[width=0.48\textwidth]{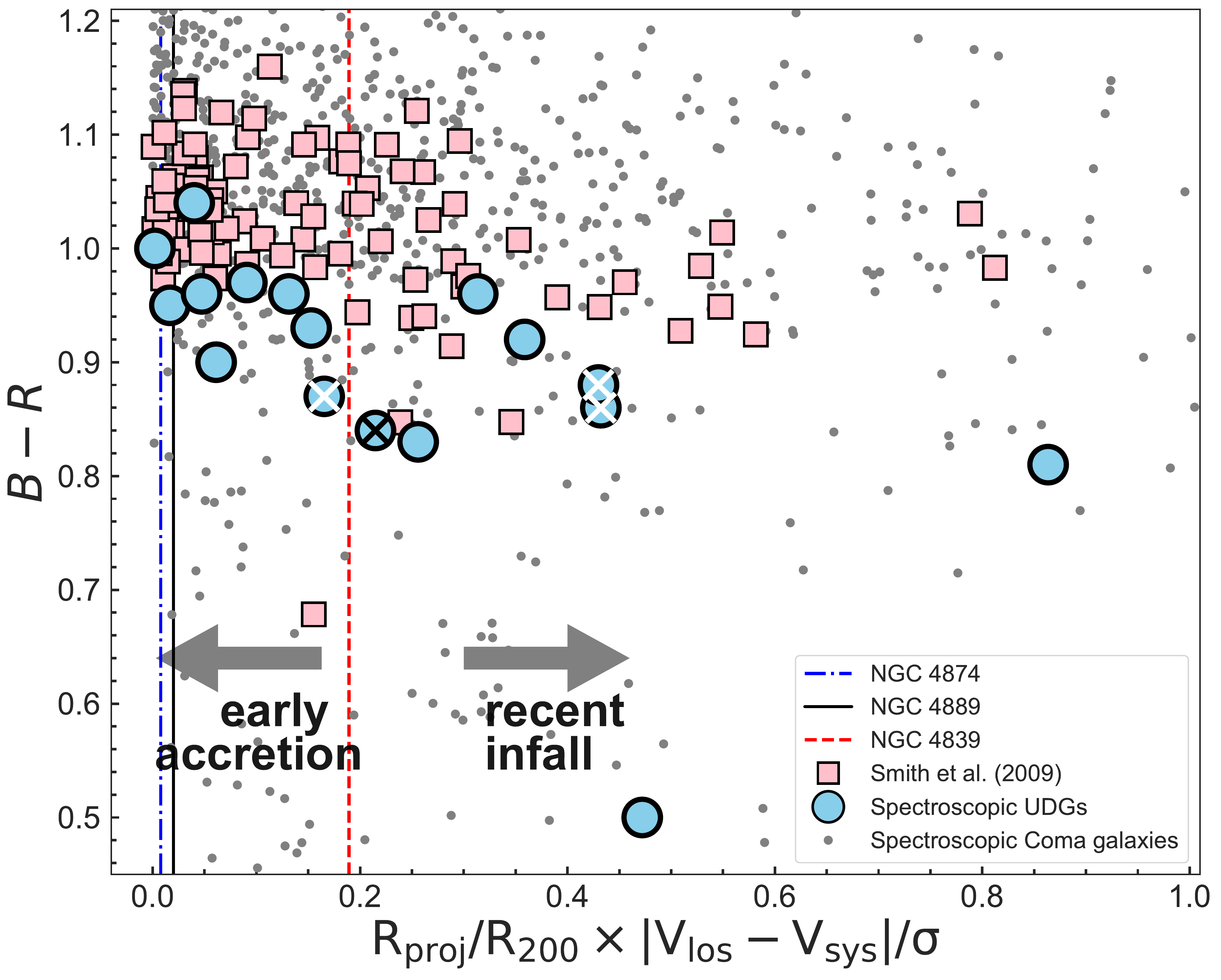}\hspace{0.01\textwidth}
	\caption{\label{fig:infall_epoch} Modified phase-space diagram showing optical color of Coma galaxies as a function of their qualitative infall time. The horizontal axis is a proxy for the infall time of galaxies into the Coma cluster. The vertical dashed lines represent the infall times of the 3 brightest cluster galaxies as shown in the plot legend. We have also included the high surface brightness dwarf galaxies from \citet{Smith_2009} in the diagram for comparison. As in previous Figures, the ultra-diffuse galaxies with inferred optical $B-R$ colors (from \citealt{Meng_2017}) have been marked with white crosses, while the low surface brightness galaxy is marked with a black cross. }
\end{figure}

To properly understand the origin of UDGs relative to other galaxy populations within the Coma cluster, we combine their spatial, velocity and optical colors information in the modified phase-space plot shown in Figure \ref{fig:infall_epoch}. The horizontal axis, $R_{\rm proj}/R_{\rm 200} \times \lvert V_{\rm los}-V_{\rm sys} \rvert/\sigma$, is now a proxy for the accretion epoch (see \citealt{Haines_2012, Noble_2013, Rhee_2017}, etc.). For easy referencing, hereafter, we will refer to this quantity as \textit{infall time} though we note that it is a dimensionless quantity. The horizontal axis is a proxy for accretion epoch because in velocity phase-space $R_{\rm proj} \times V_{\rm los}$ is constant along ``chevron''-shaped caustics, and may be used to identify infalling or virialized groups within the cluster. In Figure \ref{fig:infall_epoch}, galaxies along any vertical line share similar accretion epochs and as such, we can qualitatively infer the infall times of our UDG sample relative to the 3 brightest galaxies within Coma cluster. Galaxies with smaller $R_{\rm proj} \times V_{\rm los}$ have been in the cluster longer, and vice versa, such that NGC~4874 is at the center of the cluster potential and NGC~4839 is a relatively recent infall. According to the cosmological simulations of \citet{Rhee_2017}, the recent and ancient infalls are expected, on average, to have been in the cluster for ${\sim}2$ and ${\sim}8$~Gyr, respectively.

Both the HSB dwarf sample and our UDGs show a transition in optical colors at \textit{infall time} ${\sim0.3}$ as shown in Figure \ref{fig:infall_epoch}. This corresponds to a clustercentric radii of ${\sim}0.6$~Mpc, i.e., the cluster core boundary. While the HSB dwarfs simply become bluer and younger \citep{Smith_2009} beyond this transition radius, UDGs appear in two color families: a red sequence within the cluster core, which becomes bluer with clustercentric radii, and a blue cloud, which dominates in the recent infall region (cluster outskirts). 

From Figure \ref{fig:infall_epoch}, the recent infalls are the least red UDGs, typically characterized by high absolute line-of-sight velocities, and generally smaller. We emphasize that most of these galaxies are not co-spatial with NGC~4839 despite their similar accretion epochs. There is thus evidence that the NGC~4839 subgroup may not be the sole driver of the trends we observed earlier in the cluster outskirts, although there is a noticeable overdensity of UDGs and dwarfs with accretion epochs similar to NGC~4839. It is also clear that not all UDGs are recent infalls. A few UDGs (Yagi285, Yagi364, DF08, Yagi093)\footnote{unfortunately, we do not yet have the stellar populations of these ``primordial'' UDGs apart from Yagi093, which is ${\sim}8$~Gyr old (Paper II) and may be undergoing some disruption} are ``primordial'' in the sense that they have been in the cluster since early epochs, with accretion epochs comparable to NGC~4874 and NGC~4889. The detection of weak [O II] and [O III] emission lines was reported in \citet{Kadowaki_2017} for DF08, although our analysis here suggests it is an ancient infall into the cluster. This could imply that DF08 is massive enough to have retained its gas for ${\sim}8$~Gyr. While we do not yet have optical color information for this peculiar UDG, we envisage that it should be similarly red like other ancient infalls we have identified in this study, and note that it is a prime candidate for future stellar population studies. The low absolute line-of-sight velocities of these ancient infalls reflect the mass of the cluster when they were accreted (or probably formed in-situ). A caveat that should be borne in mind when interpreting our modified phase-space diagram is the confusing effect of projection which could make the \textit{infall time} we report here lower limits.

\textit{The overall picture that emerges from these results for present-day cluster UDGs is one where their progenitors are accreted from the field where they may have experienced some pre-processing. These progenitors, which are significantly bluer, acquire radial velocities corresponding to the mass of the cluster at accretion as they are accelerated towards the cluster center. During this virialization process, they experience star-formation quenching (assuming they are not already quiescent), become redder and bigger, and they may also be morphologically transformed due to the various physical processes operating within the cluster}.

\section{Summary and prospects}
In this work we have spectroscopically confirmed the cluster membership of 16 new ultra-diffuse galaxies within the Coma cluster using Keck/DEIMOS data. This brings the total number of UDGs within the Coma cluster with line-of-sight velocities to 24. With this modest kinematics sample, we have performed a pilot exploration of UDGs in velocity phase-space with the aim of understanding their origins. We find evidence that:
\begin{itemize}
\item the kinematics of ultra-diffuse galaxies are consistent with other galaxies in their vicinity, even though our ultra-diffuse galaxy sample is not as relaxed as the red galaxy subpopulation within the cluster,
\item at any clustercentric radii within the cluster core, ultra-diffuse galaxies with higher absolute velocities have bluer optical colors, unlike their slowly moving counterparts which are redder,
\item most ultra-diffuse galaxies in our sample show evidence of recent infall into the cluster while some are primordial in the sense that they share similar accretion epochs with the most luminous galaxies within the cluster,
\end{itemize}  

The results presented in this pilot work show that while UDGs may or may not be spatially associated with visible kinematic substructures in cluster environments, their velocity distribution, which is similar to that of the late-type cluster galaxies, show signs that are compatible with an infall origin of progenitors, over multiple episodes of accretion. As more data become available from different parts of the cluster, it would be interesting to see how Figure \ref{fig:infall_epoch} evolves. Deep spectroscopy of UDGs from the Coma cluster outskirts as well as in the field environments is definitely needed to properly understand their true origin. Lastly, it may now be possible to identify infalling galaxies that are being transformed into ultra-diffuse galaxies in the modified phase-space diagram.

\section*{Acknowledgements}
We wish to thank Pieter van Dokkum for providing the D16 DEIMOS mask data.
This work was supported by NSF grants AST-1211995 and AST-1616598.
AJR was supported by NSF grant AST-1515084 and as a Research Corporation for Science Advancement Cottrell Scholar.
AFM and DAF thank the ARC for financial support via DP160101608. SB acknowledges the support of the AAO PhD Topup Scholarship.

\bibliographystyle{mnras}
\bibliography{coma_kin}

\begin{thebibliography}{}
\makeatletter
\relax
\def\mn@urlcharsother{\let\do\@makeother \do\$\do\&\do\#\do\^\do\_\do\%\do\~}
\def\mn@doi{\begingroup\mn@urlcharsother \@ifnextchar [ {\mn@doi@}
  {\mn@doi@[]}}
\def\mn@doi@[#1]#2{\def\@tempa{#1}\ifx\@tempa\@empty \href
  {http://dx.doi.org/#2} {doi:#2}\else \href {http://dx.doi.org/#2} {#1}\fi
  \endgroup}
\def\mn@eprint#1#2{\mn@eprint@#1:#2::\@nil}
\def\mn@eprint@arXiv#1{\href {http://arxiv.org/abs/#1} {{\tt arXiv:#1}}}
\def\mn@eprint@dblp#1{\href {http://dblp.uni-trier.de/rec/bibtex/#1.xml}
  {dblp:#1}}
\def\mn@eprint@#1:#2:#3:#4\@nil{\def\@tempa {#1}\def\@tempb {#2}\def\@tempc
  {#3}\ifx \@tempc \@empty \let \@tempc \@tempb \let \@tempb \@tempa \fi \ifx
  \@tempb \@empty \def\@tempb {arXiv}\fi \@ifundefined
  {mn@eprint@\@tempb}{\@tempb:\@tempc}{\expandafter \expandafter \csname
  mn@eprint@\@tempb\endcsname \expandafter{\@tempc}}}

\bibitem[\protect\citeauthoryear{{Abolfathi} et~al.,}{{Abolfathi}
  et~al.}{2017}]{SDSS}
{Abolfathi} B.,  et~al., 2017, preprint, \href
  {http://adsabs.harvard.edu/abs/2017arXiv170709322A} {} (\mn@eprint {arXiv}
  {1707.09322})

\bibitem[\protect\citeauthoryear{{Adami}, {Biviano}, {Durret}  \&
  {Mazure}}{{Adami} et~al.}{2005}]{Adami_2005}
{Adami} C.,  {Biviano} A.,  {Durret} F.,   {Mazure} A.,  2005, \mn@doi [\aap]
  {10.1051/0004-6361:20053504}, \href
  {http://adsabs.harvard.edu/abs/2005A%26A...443...17A} {443, 17}

\bibitem[\protect\citeauthoryear{{Adami} et~al.,}{{Adami}
  et~al.}{2006}]{Adami_2006}
{Adami} C.,  et~al., 2006, \mn@doi [\aap] {10.1051/0004-6361:20053810}, \href
  {http://adsabs.harvard.edu/abs/2006A%26A...451.1159A} {451, 1159}

\bibitem[\protect\citeauthoryear{{Amorisco}, {Monachesi}, {Agnello}  \&
  {White}}{{Amorisco} et~al.}{2016}]{Amorisco_2016}
{Amorisco} N.~C.,  {Monachesi} A.,  {Agnello} A.,   {White} S.~D.~M.,  2016,
  preprint, \href {http://adsabs.harvard.edu/abs/2016arXiv161001595A} {}
  (\mn@eprint {arXiv} {1610.01595})

\bibitem[\protect\citeauthoryear{{Beasley} \& {Trujillo}}{{Beasley} \&
  {Trujillo}}{2016}]{Beasley_2016}
{Beasley} M.~A.,  {Trujillo} I.,  2016, \mn@doi [\apj]
  {10.3847/0004-637X/830/1/23}, \href
  {http://adsabs.harvard.edu/abs/2016ApJ...830...23B} {830, 23}

\bibitem[\protect\citeauthoryear{{Bertschinger}}{{Bertschinger}}{1985}]{Bertschinger_1985}
{Bertschinger} E.,  1985, \mn@doi [\apjs] {10.1086/191028}, \href
  {http://adsabs.harvard.edu/abs/1985ApJS...58...39B} {58, 39}

\bibitem[\protect\citeauthoryear{{Binney} \& {Tremaine}}{{Binney} \&
  {Tremaine}}{2008}]{Binney_2008}
{Binney} J.,  {Tremaine} S.,  2008, {Galactic Dynamics: Second Edition}.
Princeton University Press

\bibitem[\protect\citeauthoryear{{Biviano}, {Durret}, {Gerbal}, {Le Fevre},
  {Lobo}, {Mazure}  \& {Slezak}}{{Biviano} et~al.}{1996}]{Biviano_1996}
{Biviano} A.,  {Durret} F.,  {Gerbal} D.,  {Le Fevre} O.,  {Lobo} C.,  {Mazure}
  A.,   {Slezak} E.,  1996, \aap, \href
  {http://adsabs.harvard.edu/abs/1996A%26A...311...95B} {311, 95}

\bibitem[\protect\citeauthoryear{{Boselli} \& {Gavazzi}}{{Boselli} \&
  {Gavazzi}}{2006}]{Boselli_2006}
{Boselli} A.,  {Gavazzi} G.,  2006, \mn@doi [\pasp] {10.1086/500691}, \href
  {http://adsabs.harvard.edu/abs/2006PASP..118..517B} {118, 517}

\bibitem[\protect\citeauthoryear{{Briel} et~al.,}{{Briel}
  et~al.}{2001}]{Briel_2001}
{Briel} U.~G.,  et~al., 2001, \mn@doi [\aap] {10.1051/0004-6361:20000024},
  \href {http://adsabs.harvard.edu/abs/2001A%26A...365L..60B} {365, L60}

\bibitem[\protect\citeauthoryear{{Burkert}}{{Burkert}}{2017}]{Burkert_2017}
{Burkert} A.,  2017, \mn@doi [\apj] {10.3847/1538-4357/aa671c}, \href
  {http://adsabs.harvard.edu/abs/2017ApJ...838...93B} {838, 93}

\bibitem[\protect\citeauthoryear{{Cardiel}}{{Cardiel}}{1999}]{Cardiel_1999}
{Cardiel} N.,  1999, PhD thesis, , Universidad Complutense de Madrid, Spain,
  (1999)

\bibitem[\protect\citeauthoryear{{Chan}, {Kere{\v s}}, {Wetzel}, {Hopkins},
  {Faucher-Gigu{\`e}re}, {El-Badry}, {Garrison-Kimmel}  \&
  {Boylan-Kolchin}}{{Chan} et~al.}{2017}]{Chan_2017}
{Chan} T.~K.,  {Kere{\v s}} D.,  {Wetzel} A.,  {Hopkins} P.~F.,
  {Faucher-Gigu{\`e}re} C.-A.,  {El-Badry} K.,  {Garrison-Kimmel} S.,
  {Boylan-Kolchin} M.,  2017, preprint, \href
  {http://adsabs.harvard.edu/abs/2017arXiv171104788C} {} (\mn@eprint {arXiv}
  {1711.04788})

\bibitem[\protect\citeauthoryear{{Colless} \& {Dunn}}{{Colless} \&
  {Dunn}}{1996}]{Colless_1996}
{Colless} M.,  {Dunn} A.~M.,  1996, \mn@doi [\apj] {10.1086/176827}, \href
  {http://adsabs.harvard.edu/abs/1996ApJ...458..435C} {458, 435}

\bibitem[\protect\citeauthoryear{{Conselice}, {Gallagher}  \&
  {Wyse}}{{Conselice} et~al.}{2001}]{Conselice_2001}
{Conselice} C.~J.,  {Gallagher} III J.~S.,   {Wyse} R.~F.~G.,  2001, \mn@doi
  [\apj] {10.1086/322373}, \href
  {http://adsabs.harvard.edu/abs/2001ApJ...559..791C} {559, 791}

\bibitem[\protect\citeauthoryear{{Cooper}, {Newman}, {Davis}, {Finkbeiner}  \&
  {Gerke}}{{Cooper} et~al.}{2012}]{Cooper_2012}
{Cooper} M.~C.,  {Newman} J.~A.,  {Davis} M.,  {Finkbeiner} D.~P.,   {Gerke}
  B.~F.,  2012, {spec2d: DEEP2 DEIMOS Spectral Pipeline}, Astrophysics Source
  Code Library (\mn@eprint {ascl} {1203.003})

\bibitem[\protect\citeauthoryear{{Di Cintio}, {Brook}, {Dutton}, {Macci{\`o}},
  {Obreja}  \& {Dekel}}{{Di Cintio} et~al.}{2017}]{DiCintio_2017}
{Di Cintio} A.,  {Brook} C.~B.,  {Dutton} A.~A.,  {Macci{\`o}} A.~V.,  {Obreja}
  A.,   {Dekel} A.,  2017, \mn@doi [\mnras] {10.1093/mnrasl/slw210}, \href
  {http://adsabs.harvard.edu/abs/2017MNRAS.466L...1D} {466, L1}

\bibitem[\protect\citeauthoryear{{Godwin}, {Metcalfe}  \& {Peach}}{{Godwin}
  et~al.}{1983}]{Godwin_1983}
{Godwin} J.~G.,  {Metcalfe} N.,   {Peach} J.~V.,  1983, \mn@doi [\mnras]
  {10.1093/mnras/202.1.113}, \href
  {http://adsabs.harvard.edu/abs/1983MNRAS.202..113G} {202, 113}

\bibitem[\protect\citeauthoryear{{Gu} et~al.,}{{Gu} et~al.}{2017}]{Meng_2017}
{Gu} M.,  et~al., 2017, preprint, \href
  {http://adsabs.harvard.edu/abs/2017arXiv170907003G} {} (\mn@eprint {arXiv}
  {1709.07003})

\bibitem[\protect\citeauthoryear{{Haines} et~al.,}{{Haines}
  et~al.}{2012}]{Haines_2012}
{Haines} C.~P.,  et~al., 2012, \mn@doi [\apj] {10.1088/0004-637X/754/2/97},
  \href {http://adsabs.harvard.edu/abs/2012ApJ...754...97H} {754, 97}

\bibitem[\protect\citeauthoryear{{Haines} et~al.,}{{Haines}
  et~al.}{2015}]{Haines_2015}
{Haines} C.~P.,  et~al., 2015, \mn@doi [\apj] {10.1088/0004-637X/806/1/101},
  \href {http://adsabs.harvard.edu/abs/2015ApJ...806..101H} {806, 101}

\bibitem[\protect\citeauthoryear{{Janssens}, {Abraham}, {Brodie}, {Forbes},
  {Romanowsky}  \& {van Dokkum}}{{Janssens} et~al.}{2017}]{Janssens_2017}
{Janssens} S.,  {Abraham} R.,  {Brodie} J.,  {Forbes} D.,  {Romanowsky} A.~J.,
   {van Dokkum} P.,  2017, \mn@doi [\apjl] {10.3847/2041-8213/aa667d}, \href
  {http://adsabs.harvard.edu/abs/2017ApJ...839L..17J} {839, L17}

\bibitem[\protect\citeauthoryear{{Kadowaki}, {Zaritsky}  \&
  {Donnerstein}}{{Kadowaki} et~al.}{2017}]{Kadowaki_2017}
{Kadowaki} J.,  {Zaritsky} D.,   {Donnerstein} R.~L.,  2017, \mn@doi [\apjl]
  {10.3847/2041-8213/aa653d}, \href
  {http://adsabs.harvard.edu/abs/2017ApJ...838L..21K} {838, L21}

\bibitem[\protect\citeauthoryear{{Kubo}, {Stebbins}, {Annis}, {Dell'Antonio},
  {Lin}, {Khiabanian}  \& {Frieman}}{{Kubo} et~al.}{2007}]{Kubo_2007}
{Kubo} J.~M.,  {Stebbins} A.,  {Annis} J.,  {Dell'Antonio} I.~P.,  {Lin} H.,
  {Khiabanian} H.,   {Frieman} J.~A.,  2007, \mn@doi [\apj] {10.1086/523101},
  \href {http://adsabs.harvard.edu/abs/2007ApJ...671.1466K} {671, 1466}

\bibitem[\protect\citeauthoryear{{Mahajan}, {Mamon}  \&
  {Raychaudhury}}{{Mahajan} et~al.}{2011}]{Mahajan_2011}
{Mahajan} S.,  {Mamon} G.~A.,   {Raychaudhury} S.,  2011, \mn@doi [\mnras]
  {10.1111/j.1365-2966.2011.19236.x}, \href
  {http://adsabs.harvard.edu/abs/2011MNRAS.416.2882M} {416, 2882}

\bibitem[\protect\citeauthoryear{{Makarov}, {Prugniel}, {Terekhova}, {Courtois}
   \& {Vauglin}}{{Makarov} et~al.}{2014}]{Hyperleda}
{Makarov} D.,  {Prugniel} P.,  {Terekhova} N.,  {Courtois} H.,   {Vauglin} I.,
  2014, \mn@doi [\aap] {10.1051/0004-6361/201423496}, \href
  {http://adsabs.harvard.edu/abs/2014A%26A...570A..13M} {570, A13}

\bibitem[\protect\citeauthoryear{{Mamon}, {Sanchis}, {Salvador-Sol{\'e}}  \&
  {Solanes}}{{Mamon} et~al.}{2004}]{Mamon_2004}
{Mamon} G.~A.,  {Sanchis} T.,  {Salvador-Sol{\'e}} E.,   {Solanes} J.~M.,
  2004, \mn@doi [\aap] {10.1051/0004-6361:20034155}, \href
  {http://adsabs.harvard.edu/abs/2004A%26A...414..445M} {414, 445}

\bibitem[\protect\citeauthoryear{{Mart{\'{\i}}nez-Delgado}
  et~al.,}{{Mart{\'{\i}}nez-Delgado} et~al.}{2016}]{Martinez_2016}
{Mart{\'{\i}}nez-Delgado} D.,  et~al., 2016, \mn@doi [\aj]
  {10.3847/0004-6256/151/4/96}, \href
  {http://adsabs.harvard.edu/abs/2016AJ....151...96M} {151, 96}

\bibitem[\protect\citeauthoryear{{Mendelin} \& {Binggeli}}{{Mendelin} \&
  {Binggeli}}{2017}]{Mendelin_2017}
{Mendelin} M.,  {Binggeli} B.,  2017, \mn@doi [\aap]
  {10.1051/0004-6361/201730567}, \href
  {http://adsabs.harvard.edu/abs/2017A%26A...604A..96M} {604, A96}

\bibitem[\protect\citeauthoryear{{Neumann}, {Lumb}, {Pratt}  \&
  {Briel}}{{Neumann} et~al.}{2003}]{Neumann_2003}
{Neumann} D.~M.,  {Lumb} D.~H.,  {Pratt} G.~W.,   {Briel} U.~G.,  2003, \mn@doi
  [\aap] {10.1051/0004-6361:20021911}, \href
  {http://adsabs.harvard.edu/abs/2003A%26A...400..811N} {400, 811}

\bibitem[\protect\citeauthoryear{{Noble}, {Webb}, {Muzzin}, {Wilson}, {Yee}  \&
  {van der Burg}}{{Noble} et~al.}{2013}]{Noble_2013}
{Noble} A.~G.,  {Webb} T.~M.~A.,  {Muzzin} A.,  {Wilson} G.,  {Yee} H.~K.~C.,
  {van der Burg} R.~F.~J.,  2013, \mn@doi [\apj] {10.1088/0004-637X/768/2/118},
  \href {http://adsabs.harvard.edu/abs/2013ApJ...768..118N} {768, 118}

\bibitem[\protect\citeauthoryear{{Oman}, {Hudson}  \& {Behroozi}}{{Oman}
  et~al.}{2013}]{Oman_2013}
{Oman} K.~A.,  {Hudson} M.~J.,   {Behroozi} P.~S.,  2013, \mn@doi [\mnras]
  {10.1093/mnras/stt328}, \href
  {http://adsabs.harvard.edu/abs/2013MNRAS.431.2307O} {431, 2307}

\bibitem[\protect\citeauthoryear{{Pandya} et~al.,}{{Pandya}
  et~al.}{2017}]{Viraj_2017}
{Pandya} V.,  et~al., 2017, preprint, \href
  {http://adsabs.harvard.edu/abs/2017arXiv171105272P} {} (\mn@eprint {arXiv}
  {1711.05272})

\bibitem[\protect\citeauthoryear{{Rhee}, {Smith}, {Choi}, {Yi}, {Jaff{\'e}},
  {Candlish}  \& {S{\'a}nchez-J{\'a}nssen}}{{Rhee} et~al.}{2017}]{Rhee_2017}
{Rhee} J.,  {Smith} R.,  {Choi} H.,  {Yi} S.~K.,  {Jaff{\'e}} Y.,  {Candlish}
  G.,   {S{\'a}nchez-J{\'a}nssen} R.,  2017, \mn@doi [\apj]
  {10.3847/1538-4357/aa6d6c}, \href
  {http://adsabs.harvard.edu/abs/2017ApJ...843..128R} {843, 128}

\bibitem[\protect\citeauthoryear{{Rom{\'a}n} \& {Trujillo}}{{Rom{\'a}n} \&
  {Trujillo}}{2017a}]{Roman_2017a}
{Rom{\'a}n} J.,  {Trujillo} I.,  2017a, \mn@doi [\mnras]
  {10.1093/mnras/stx438}, \href
  {http://adsabs.harvard.edu/abs/2017MNRAS.468..703R} {468, 703}

\bibitem[\protect\citeauthoryear{{Rom{\'a}n} \& {Trujillo}}{{Rom{\'a}n} \&
  {Trujillo}}{2017b}]{Roman_2017b}
{Rom{\'a}n} J.,  {Trujillo} I.,  2017b, \mn@doi [\mnras]
  {10.1093/mnras/stx694}, \href
  {http://adsabs.harvard.edu/abs/2017MNRAS.468.4039R} {468, 4039}

\bibitem[\protect\citeauthoryear{{Rong}, {Guo}, {Gao}, {Liao}, {Xie}, {Puzia},
  {Sun}  \& {Pan}}{{Rong} et~al.}{2017}]{Rong_2017}
{Rong} Y.,  {Guo} Q.,  {Gao} L.,  {Liao} S.,  {Xie} L.,  {Puzia} T.~H.,  {Sun}
  S.,   {Pan} J.,  2017, \mn@doi [\mnras] {10.1093/mnras/stx1440}, \href
  {http://adsabs.harvard.edu/abs/2017MNRAS.470.4231R} {470, 4231}

\bibitem[\protect\citeauthoryear{{Sarazin}}{{Sarazin}}{1986}]{Sarazin_1986}
{Sarazin} C.~L.,  1986, \mn@doi [Reviews of Modern Physics]
  {10.1103/RevModPhys.58.1}, \href
  {http://adsabs.harvard.edu/abs/1986RvMP...58....1S} {58, 1}

\bibitem[\protect\citeauthoryear{{Shi} et~al.,}{{Shi} et~al.}{2017}]{Shi_2017}
{Shi} D.~D.,  et~al., 2017, \mn@doi [\apj] {10.3847/1538-4357/aa8327}, \href
  {http://adsabs.harvard.edu/abs/2017ApJ...846...26S} {846, 26}

\bibitem[\protect\citeauthoryear{{Smith}, {Lucey}, {Hudson}, {Allanson},
  {Bridges}, {Hornschemeier}, {Marzke}  \& {Miller}}{{Smith}
  et~al.}{2009}]{Smith_2009}
{Smith} R.~J.,  {Lucey} J.~R.,  {Hudson} M.~J.,  {Allanson} S.~P.,  {Bridges}
  T.~J.,  {Hornschemeier} A.~E.,  {Marzke} R.~O.,   {Miller} N.~A.,  2009,
  \mn@doi [\mnras] {10.1111/j.1365-2966.2008.14180.x}, \href
  {http://adsabs.harvard.edu/abs/2009MNRAS.392.1265S} {392, 1265}

\bibitem[\protect\citeauthoryear{{Smith}, {Lucey}, {Price}, {Hudson}  \&
  {Phillipps}}{{Smith} et~al.}{2012}]{Smith_2012}
{Smith} R.~J.,  {Lucey} J.~R.,  {Price} J.,  {Hudson} M.~J.,   {Phillipps} S.,
  2012, \mn@doi [\mnras] {10.1111/j.1365-2966.2011.19956.x}, \href
  {http://adsabs.harvard.edu/abs/2012MNRAS.419.3167S} {419, 3167}

\bibitem[\protect\citeauthoryear{{Spekkens} \& {Karunakaran}}{{Spekkens} \&
  {Karunakaran}}{2017}]{Spekkens_2017}
{Spekkens} K.,  {Karunakaran} A.,  2017, preprint, \href
  {http://adsabs.harvard.edu/abs/2017arXiv171006557S} {} (\mn@eprint {arXiv}
  {1710.06557})

\bibitem[\protect\citeauthoryear{{Tonry} \& {Davis}}{{Tonry} \&
  {Davis}}{1979}]{Tonry_1979}
{Tonry} J.,  {Davis} M.,  1979, \mn@doi [\aj] {10.1086/112569}, \href
  {http://adsabs.harvard.edu/abs/1979AJ.....84.1511T} {84, 1511}

\bibitem[\protect\citeauthoryear{{Vazdekis}, {Koleva}, {Ricciardelli},
  {R{\"o}ck}  \& {Falc{\'o}n-Barroso}}{{Vazdekis} et~al.}{2016}]{Vazdekis_2016}
{Vazdekis} A.,  {Koleva} M.,  {Ricciardelli} E.,  {R{\"o}ck} B.,
  {Falc{\'o}n-Barroso} J.,  2016, \mn@doi [\mnras] {10.1093/mnras/stw2231},
  \href {http://adsabs.harvard.edu/abs/2016MNRAS.463.3409V} {463, 3409}

\bibitem[\protect\citeauthoryear{{Vijayaraghavan}, {Gallagher}  \&
  {Ricker}}{{Vijayaraghavan} et~al.}{2015}]{VijarX_2015}
{Vijayaraghavan} R.,  {Gallagher} J.~S.,   {Ricker} P.~M.,  2015, \mn@doi
  [\mnras] {10.1093/mnras/stu2761}, \href
  {http://adsabs.harvard.edu/abs/2015MNRAS.447.3623V} {447, 3623}

\bibitem[\protect\citeauthoryear{{Yagi}, {Yamanoi}, {Furusawa}, {Nakata}  \&
  {Komiyama}}{{Yagi} et~al.}{2013}]{Yagi_2013}
{Yagi} Nao M.~S.,  {Yamanoi} H.,  {Furusawa} H.,  {Nakata} F.,   {Komiyama} Y.,
   2013, \mn@doi [\pasj] {10.1093/pasj/65.1.22}, \href
  {http://adsabs.harvard.edu/abs/2013PASJ...65...22Y} {65, 22}

\bibitem[\protect\citeauthoryear{{Yagi}, {Koda}, {Komiyama}  \&
  {Yamanoi}}{{Yagi} et~al.}{2016}]{Yagi_2016}
{Yagi} M.,  {Koda} J.,  {Komiyama} Y.,   {Yamanoi} H.,  2016, \mn@doi [\apjs]
  {10.3847/0067-0049/225/1/11}, \href
  {http://adsabs.harvard.edu/abs/2016ApJS..225...11Y} {225, 11}

\bibitem[\protect\citeauthoryear{{Yamanoi} et~al.,}{{Yamanoi}
  et~al.}{2012}]{Yamanoi_2012}
{Yamanoi} H.,  et~al., 2012, \mn@doi [\aj] {10.1088/0004-6256/144/2/40}, \href
  {http://adsabs.harvard.edu/abs/2012AJ....144...40Y} {144, 40}

\bibitem[\protect\citeauthoryear{{van Dokkum} et~al.,}{{van Dokkum}
  et~al.}{2015}]{vanDokkum_2015}
{van Dokkum} P.~G.,  et~al., 2015, \mn@doi [\apjl]
  {10.1088/2041-8205/804/1/L26}, \href
  {http://adsabs.harvard.edu/abs/2015ApJ...804L..26V} {804, L26}

\bibitem[\protect\citeauthoryear{{van Dokkum} et~al.,}{{van Dokkum}
  et~al.}{2016}]{vanDokkum_2016}
{van Dokkum} P.,  et~al., 2016, \mn@doi [\apjl] {10.3847/2041-8205/828/1/L6},
  \href {http://adsabs.harvard.edu/abs/2016ApJ...828L...6V} {828, L6}

\bibitem[\protect\citeauthoryear{{van Dokkum} et~al.,}{{van Dokkum}
  et~al.}{2017}]{vanDokkum_2017}
{van Dokkum} P.,  et~al., 2017, \mn@doi [\apjl] {10.3847/2041-8213/aa7ca2},
  \href {http://adsabs.harvard.edu/abs/2017ApJ...844L..11V} {844, L11}

\bibitem[\protect\citeauthoryear{{van der Burg}, {Muzzin}  \& {Hoekstra}}{{van
  der Burg} et~al.}{2016}]{vdBurg_2016}
{van der Burg} R.~F.~J.,  {Muzzin} A.,   {Hoekstra} H.,  2016, \mn@doi [\aap]
  {10.1051/0004-6361/201628222}, \href
  {http://adsabs.harvard.edu/abs/2016A%26A...590A..20V} {590, A20}

\bibitem[\protect\citeauthoryear{{van der Burg} et~al.,}{{van der Burg}
  et~al.}{2017}]{vdBurg_2017}
{van der Burg} R.~F.~J.,  et~al., 2017, \mn@doi [\aap]
  {10.1051/0004-6361/201731335}, \href
  {http://adsabs.harvard.edu/abs/2017A%26A...607A..79V} {607, A79}

\makeatother
\end{thebibliography}

\appendix

\bsp

\end{document}